\documentclass[journal]{vgtc}                
\ifpdf
  \pdfoutput=1\relax                   
  \usepackage{graphicx}                
  \DeclareGraphicsExtensions{.pdf,.png,.jpg,.jpeg} 
\else
  \usepackage{graphicx}                
  \DeclareGraphicsExtensions{.eps}     
\fi%

\graphicspath{{figures/}{pictures/}{images/}{./}} 

\usepackage{microtype}                 
\PassOptionsToPackage{warn}{textcomp}  
\usepackage{textcomp}                  
\usepackage{mathptmx}                  
\usepackage{times}                     
\usepackage{cite}                      
\usepackage{tabu}                      
\usepackage{booktabs}                  
\usepackage{amsmath,amssymb}
\usepackage{mathtools} 
\usepackage{ltl}
\usepackage{layouts}
\usepackage{scalerel}
\usepackage{siunitx}
\usepackage{blindtext}
\usepackage{tikz}

\onlineid{0}

\vgtccategory{Research}
\vgtcpapertype{application/design study}

\title{Visual Analysis of Hyperproperties\\ for Understanding Model Checking Results}
\author{Tom Horak, Norine Coenen, Niklas Metzger, Christopher Hahn, Tamara Flemisch,\\ Juli\'an M\'endez, Dennis Dimov, Bernd Finkbeiner, and Raimund Dachselt}
\authorfooter{
\item
 T. Horak, T. Flemisch, J. M\'endez, and D. Dimov are with the Interactive Media Lab at Technische Universit\"{a}t Dresden. Emails: horakt@acm.org, tamara.flemisch@tu-dresden.de, dennis.dimov@tu-dresden.de, julian\_jesus.mendez\_oconitrillo@mailbox.tu-dresden.de
\item
 R. Dachselt is with the Interactive Media Lab, the Centre for Tactile Internet (CeTI), and the Cluster of Excellence Physics of Life (PoL) at Technische Universit\"{a}t Dresden. Email: dachselt@acm.org
\item
N. Coenen, N. Metzger, C. Hahn, and B. Finkbeiner are with the Reactive Systems Group at CISPA Helmholtz Center for Information Security. Emails: \{norine.coenen, niklas.metzger, christopher.hahn, finkbeiner\}@cispa.de
}

\shortauthortitle{Horak \MakeLowercase{\textit{et al.}}: HyperVis}

\abstract{
Model checkers provide algorithms for proving that a mathematical model of a system satisfies a given specification.
In case of a violation, a counterexample that shows the erroneous behavior is returned.
Understanding these counterexamples is challenging, especially for hyperproperty specifications, i.e., specifications that relate multiple executions of a system to each other.
We aim to facilitate the visual analysis of such counterexamples through our \hypervis tool, which provides interactive visualizations of the given model, specification, and counterexample.
Within an iterative and interdisciplinary design process, we developed visualization solutions that can effectively communicate the core aspects of the model checking result.
Specifically, we introduce graphical representations of binary values for improving pattern recognition, color encoding for better indicating related aspects, visually enhanced textual descriptions, as well as extensive cross-view highlighting mechanisms.
Further, through an underlying causal analysis of the counterexample, we are also able to identify values that contributed to the violation and use this knowledge for both improved encoding and highlighting.
Finally, the analyst can modify both the specification of the hyperproperty and the system directly within \hypervis and initiate the model checking of the new version.
In combination, these features notably support the analyst in understanding the error leading to the counterexample as well as iterating the provided system and specification.
We ran multiple case studies with \hypervis and tested it with domain experts in qualitative feedback sessions. The participants' positive feedback confirms the considerable improvement over the manual, text-based status quo and the value of the tool for explaining hyperproperties.
} 


\CCScatlist{ 
 \CCScat{K.6.1}{Management of Computing and Information Systems}%
{Project and People Management}{Life Cycle};
 \CCScat{K.7.m}{The Computing Profession}{Miscellaneous}{Ethics}
}

\teaser{
  \centering
  \includegraphics[width=\linewidth]{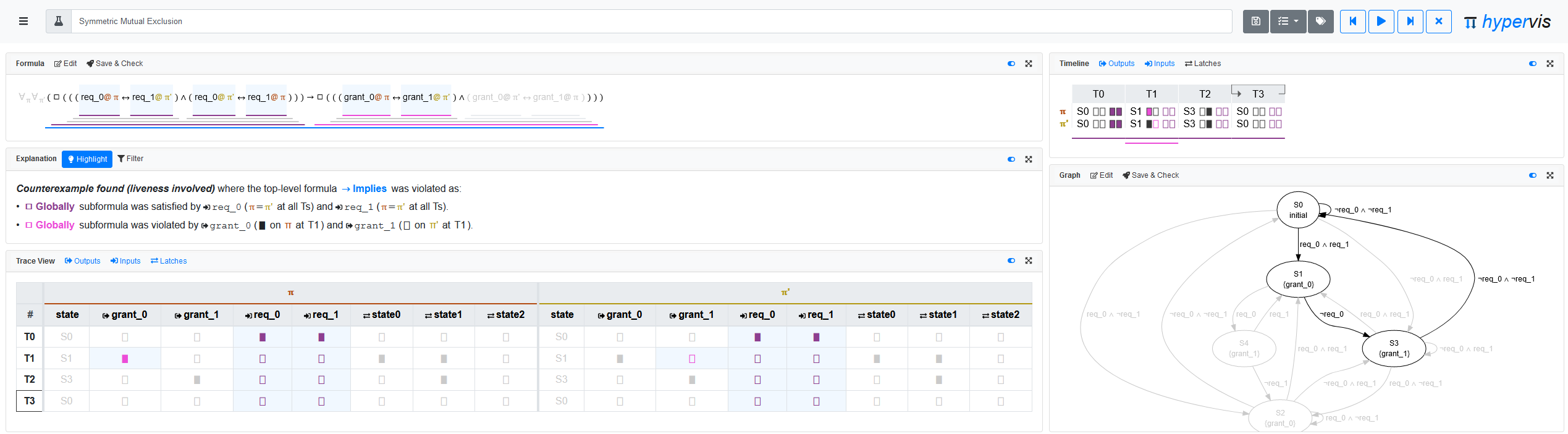}
  \vspace{-20pt}
  \caption{\hypervis provides linked interactive views with highlighting mechanisms for analyzing counterexamples of hyperproperties.}
  \label{fig:teaser}
}

\usepackage{xspace}

\newcommand*{\formulaView}{formula view\xspace}
\newcommand*{\traceView}{trace view\xspace}
\newcommand*{\graphView}{graph view\xspace}
\newcommand*{\explanationView}{explanation view\xspace}
\newcommand*{\timelineView}{timeline view\xspace}
\newcommand*{\FormulaView}{Formula View\xspace}
\newcommand*{\TraceView}{Trace View\xspace}
\newcommand*{\GraphView}{Graph View\xspace}
\newcommand*{\ExplanationView}{Explanation View\xspace}
\newcommand*{\TimelineView}{Timeline View\xspace}

\newcommand*{\tool}[1]{\textsc{#1}}
\newcommand*{\hypervis}{\tool{HyperVis}\xspace}
\newcommand*{\hyperltl}{\tool{HyperLTL}\xspace}
\newcommand*{\mchyper}{\tool{MCHyper}\xspace}
\newcommand*{\aiger}{\tool{AIGER}\xspace}
\newcommand{\abc}[0]{\tool{ABC}\xspace}

\newcommand{\true}[0]{\mathit{true}}
\newcommand{\false}[0]{\mathit{false}}

\newcommand{\U}{\LTLuntil}
\newcommand{\X}{\LTLnext}
\newcommand{\G}{\LTLglobally}
\newcommand{\F}{\LTLeventually}
\newcommand{\R}{\LTLrelease}

\newcommand*{\inlinefig}[1]{\scalerel*{\includegraphics{figures/inline-figures/#1}}{5}}
\newcommand*{\inlineTrue}{\protect\inlinefig{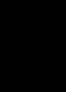}\xspace}
\newcommand*{\inlineFalse}{\protect\inlinefig{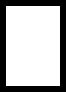}\xspace}
\newcommand*{\inlineTrueNSp}{\protect\inlinefig{inline-01}}
\newcommand*{\inlineFalseNSp}{\protect\inlinefig{inline-02}}
\newcommand*{\inlineOutput}{\protect\inlinefig{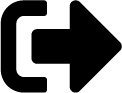}\xspace}
\newcommand*{\inlineInput}{\protect\inlinefig{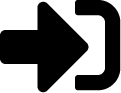}\xspace}
\newcommand*{\inlineLatch}{\protect\inlinefig{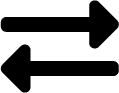}\xspace}

\definecolor{trace0}{RGB}{179, 76, 18}
\definecolor{trace1}{RGB}{179, 152, 18}
\definecolor{expl0}{RGB}{141, 63, 144}
\definecolor{expl1}{RGB}{236, 77, 216}
\definecolor{highlight}{RGB}{0, 123, 255}

\definecolor{todoRed}{RGB}{230, 40, 80}

\definecolor{revised}{RGB}{28, 68, 135}

\definecolor{href-blue}{RGB}{68, 96, 144}

\vgtcinsertpkg

\begin{document}


\firstsection{Introduction}

\maketitle


%
Model checking~\cite{10.5555/332656} is a highly efficient technique for the computer-aided verification of computer systems such as integrated circuits, network protocols, and software.
Model checking has long made the transition from research into practice and is routinely used by companies like Intel, Microsoft, or Amazon.
Intel, for example, replaced testing with verification for the core execution cluster in their design of the Intel Core i7 processor~\cite{DBLP:conf/cav/KaivolaGNTWPSTFRN09} and, recently, the initial boot code in data centers at Amazon Web Services (AWS) has been model checked to be memory safe~\cite{DBLP:conf/cav/CookKKTTT18}.
The key advantage of model checking is that it is an \emph{automatic} method: given a system description $M$ and a logical specification $\varphi$ of a desired behavioral property, the model checker automatically determines whether or not $M$ satisfies $\varphi$.
If the system design is erroneous, the model checker generates a counterexample in the form of a specific execution of $M$ that violates $\varphi$.
While finding the counterexample is completely automatic, model checking typically provides very little assistance in actually \emph{understanding} the counterexample and its underlying design flaw.
Model checkers typically output the counterexample in the form of a detailed listing that contains the complete state information for every step of a computation that leads to the violation.
Understanding all this data is already difficult for small designs and, for more complex systems and specifications, quickly becomes a daunting task.

In this paper, we present a visualization system that aids the analyst in understanding the counterexamples found by the model checker.
The visualization views communicate the core aspects of the model checking result to the analyst and support an iterative analysis process. 
We specifically focus on \emph{hyperproperties}~\cite{DBLP:journals/corr/KoleiniCM13}, a class of system specifications that is essential for the analysis of security-critical systems.
Hyperproperties express the absence of undesired dependencies or flows of information, such as those exploited in the infamous Meltdown~\cite{Lipp2018meltdown} and Spectre~\cite{Kocher2018spectre} attacks. 
A counterexample to a hyperproperty is a set of executions of the system that, together, are problematic.
For example, if some software needs to keep certain data secret, then
two executions that result from different values of the secret (but agree on the public inputs) should not show any difference on the public outputs.
Consequently, the model checker searches for such a pair of executions that differ in their public output values. 
Our goal is to help the analyst understand the violation of the hyperproperty by visualizing the relationship between the individual system executions, as well as the relationship to the system description and the logical formula. 
To support this, we have implemented the interactive tool \hypervis (\autoref{fig:teaser}, \href{https://imld.de/hypervis}{\color{href-blue}imld.de/hypervis}), that follows a multiple coordinated views approach~\cite{Roberts2007STAR_CMV, Chen2021CompositionConfigurationPatterns}.
We provide five interconnected views.
The hyperproperty specification is shown as a logical formula in the \emph{\formulaView}, the system as a state machine in the \emph{\graphView}, the executions over time in a tabular-like \emph{\traceView} and in a more compact \emph{\timelineView}. 
Additionally, there is a textual explication in the \emph{\explanationView}.

The fundamental challenge is that the connections between the different views and the relevance of their individual components is not known in advance, but rather must be deduced specifically for the hyperproperty of interest.
We address this challenge with an automated causal analysis of the counterexample, where we identify those elements of the different views that directly contribute to the violation of the specification.
The textual explication in the \emph{\explanationView} is directly based on this analysis.
In all other views, the relevant elements can be directly highlighted.
By incorporating easy-to-parse value encodings and clear color mappings alongside interactive mechanisms such as linked highlighting and debugger-like functionalities, we support the analyst in recognizing the counterexample's characteristics and in relating its different components.
Finally, after the cause of the violation is understood, the analyst can correct the system and the specification directly within the interface through integrated editing functionalities.

\hypervis is the result of an interdisciplinary effort and a highly iterative design process which included joint brainstormings and discussions between visualization and model checking experts.
The results and insights from this joint effort are presented in this paper.
Specifically, we contribute:
(1) an in-depth analysis of challenges,
(2) the design of visualization and interaction concepts enabling the visual analysis of the model checking results,
(3) the realization of these concepts with \hypervis as a web-based tool alongside integrated editing facilities, and
(4) insights from applying multiple case studies to our tool and conducting user feedback sessions with 6 participants.
In summary, our work contributes to a class of visualization solutions that aims at visually explaining complex and abstract computing concepts.

%

\begin{figure}[b]
    \vspace{-10pt}
    \centering
    \resizebox{0.75\linewidth}{!}{
    \begin{tikzpicture}
    \usetikzlibrary{calc}
    \node[align=center,draw,fill=none,minimum height=1em,minimum width=5em,rounded corners=4] (initial) {initial\\$\{\}$};
    \node[align=center,draw,fill=none,minimum height=1em,minimum width=5em,rounded corners=4] (o1) at (4,0.5) {state 1\\ $\{o_1\}$};
    \node[align=center,draw,fill=none,minimum height=1em,minimum width=5em,rounded corners=4] (o2) at (4,-0.5) {state 2\\ $\{o_2\}$};
    \node[] (init) at ($(initial.north west) + (-0.2,0.4)$){};
    \path[->,>=stealth]
    (initial) edge [line width=0.7pt, in=190,out=175,loop] node[left] {$\neg i$}(initial)
    (initial) edge [line width=0.7pt] node [above] {$i \wedge \neg s$}(o1)
    (initial) edge [line width=0.7pt] node [below] {$i \wedge s$}(o2)
    (o1) edge [line width=0.7pt,in=355,out=10,loop] node [right] {$*$} (o1)
    (o2) edge [line width=0.7pt,in=355,out=10,loop] node [right] {$*$} (o2)
    (init) edge (initial) 
    ;
  \end{tikzpicture}
}  
    
    \vspace{-7pt}
    \caption{A simple system leaking a secret~$s$ through the observable outputs~$o_1$ and~$o_2$.}
    \label{fig:info-leak}
\end{figure}
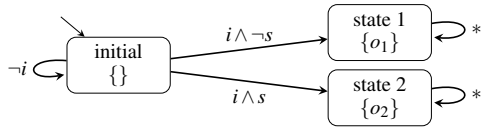

\section{Working with Hyperproperties}
\label{sec:challenges}
To support the analysis of counterexamples, we first need to understand the involved components, current workflows, and prevalent challenges.
Therefore, we will first describe the formal objects utilized during the model checking process on a toy information-flow control problem, where a system needs to satisfy observational determinism (\autoref{sec:challenges:example}).
Then, we will detail the current workflow  using a slightly bigger example (\autoref{sec:challenges:workflows}), before outlining the resulting challenges for analyzing counterexamples (\ref{sec:challenges:challenges}).

\subsection{Example: Verifying Observational Determinism}
\label{sec:challenges:example}
In general, the considered objects include the system model~$M$, the counterexample executions~$\pi$ and~$\pi'$, and the hyperproperty specification~$\varphi$.
In this simplified example, our model~$M$ (\autoref{fig:info-leak}) is prone to leak a secret $s$ via the publicly observable outputs~$o_1$ and~$o_2$ to an attacker. 
The underlying security lattice considers the secret~$s$ to be a confidential input that should not be visible to any observer while the input~$i$ and the outputs~$o_1$ and~$o_2$ are publicly observable. 
The model~$M$ can be represented as a finite state machine, where the current state determines the system's output, and the transitions of the finite state machine are labelled with the inputs to the system.
All inputs and outputs are binary values, thus, they are either present or absent.
When executing such a system, the present inputs and outputs are observed over multiple time steps.
The given system in \autoref{fig:info-leak} cycles in the first state, outputting nothing, until an input~$i$ is present. 
Depending on whether a secret~$s$ is also given, the system then either outputs~$o_2$ or~$o_1$ indefinitely.
If an attacker now happens to observe two executions of the system where the outputs are different although the input $i$ were the same on both executions, they can conclude about the secret~$s$ at this time step.

The specification $\varphi$ that we would like to verify for the system~$M$ is given as a \hyperltl formula~\cite{Clarkson2014TemporalLogicsHyperproperties}, a linear-time temporal logic for hyperproperties that can relate multiple executions.
For the example above, we would like to require observational determinism, which is formalized in \hyperltl as follows:
$
    \forall\pi ~ \forall\pi' ~ \G \left(i_\pi \leftrightarrow i_{\pi'}\right) \rightarrow \G \left(o_\pi \leftrightarrow o_{\pi'}\right).
$
The formula quantifies universally ($\forall$) over two traces $\pi$ and $\pi'$. The temporal modality~$\G$ means ``globally'', i.e., the formula $\G \varphi$~requires the subformula~$\varphi$ to hold at every point in time. 
The given formula thus states that for all trace pairs~$\pi$ and~$\pi'$ it must hold that when the observable inputs are the same at every point in time, the respective observable outputs must also be equal.
Given the model and formula, a model checker would now provide two specific executions where at a given time step the outputs differ while the inputs are equal.

\subsection{Current Workflow}
\label{sec:challenges:workflows}
\begin{figure}[b]
    \vspace{-10pt}
    \centering
    \includegraphics[width=\columnwidth]{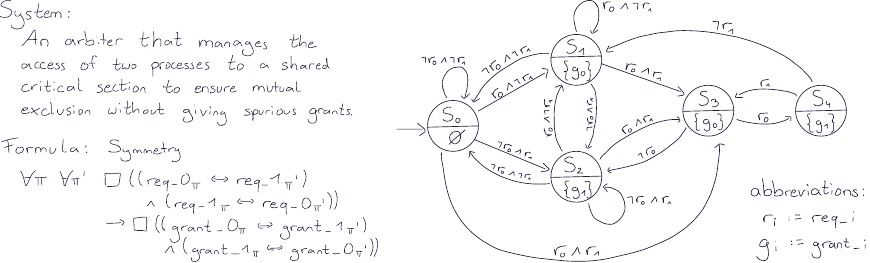}
    \vspace{-15pt}
    \caption{Handwritten details of an arbiter system and a symmetry formula checking if a process has an unfair advantage over the other process when requesting access to a shared resource.}
    \label{fig:cur-workflow-system+formula}
\end{figure}

\begin{figure}
    \centering
    \includegraphics[width=\columnwidth]{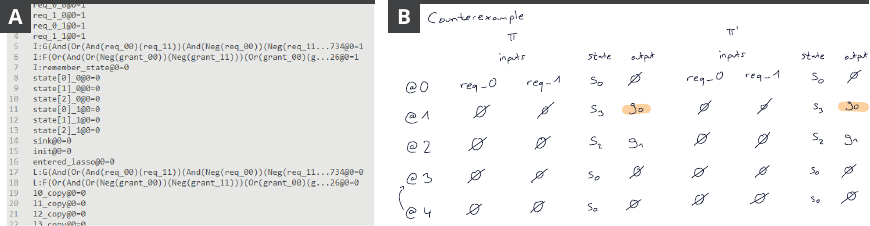}
    \vspace{-15pt}
    \caption{(a) Instance of a counterexample produced by \mchyper (excerpt), here for the system and formula described in \autoref{fig:cur-workflow-system+formula}.
    (b) Handwritten notation of the provided output in a table-like format.
    The marked outputs in the second row do not fulfil the requirements of the formula.
    }
    \label{fig:cur-workflow-cex}
    \vspace{-10pt}
\end{figure}

With the growing complexity of both the system and the specification, the model checking of hyperproperties quickly becomes complicated.
We demonstrate the current workflow and the corresponding challenges when invoking a model checker for hyperproperties on a more involved example.
To this end, we consider a system that arbitrates the access of two processes to a shared resource. 
Both processes can request access to their critical section (using $req\_i$) where they can interact with the shared resource, and the arbiter grants the access (with $grant\_i$) while ensuring mutual exclusion, i.e., only one of the processes can enter its critical section at any given time. 
The arbiter guarantees that every request will eventually be answered while not giving out spurious grants, i.e., every grant will have been requested before. 
The finite state machine for this system is sketched on the right in \autoref{fig:cur-workflow-system+formula}. 
We want to check whether the arbiter is symmetric, thus, if a pair of traces $\pi$ and $\pi'$ with symmetric requests at every step (i.e., $\G \left(req\_0_\pi \leftrightarrow req\_1_{\pi'}\right)$ and vice versa) also gives the grants symmetrically. 
This hyperproperty checks if any of the processes has an unfair advantage and is favoured when granting access to the critical section.
The corresponding \hyperltl formula expressing symmetry is noted on the left in \autoref{fig:cur-workflow-system+formula}. 
The system grants processes asymmetrically: If $\pi = \pi'$ and both processes request initially, then always process 0 is granted first (\autoref{fig:cur-workflow-cex}b).

The symmetry specification and the model of the arbiter can be given to a \hyperltl model checking tool, such as \mchyper~\cite{Finkbeiner2015AlgorithmsModelChecking}, which are typically command-line based.
The provided system models are usually considered as hardware specifications that could be implemented, e.g., on a chipset.
Consequently, the model checkers also consume low-level circuit representation like \aiger~\cite{Biere2007AIGERInverterGraph, Biere2011AIGER1.9}, encoding the system as an And-Inverter Graph. 
This representation is hard to read for human developers, who often sketch the system by hand in a more visual way (\autoref{fig:cur-workflow-system+formula}) and realize the models using hardware description languages such as \tool{Verilog}~\cite{IEEE-Verilog}, which are then compiled down to \aiger.
Given such a system description and a hyperproperty, the model checker then tries to find a counterexample, i.e., a set of system executions that together violate the \hyperltl formula.

If a violation occurred, a counterexample is reported in a textual representation where each line represents variable's value on a given trace at a given time step (\autoref{fig:cur-workflow-cex}a).
This representation is hard to grasp as even for smaller counterexamples, this output consists of a few hundred lines (140 lines for this arbiter example), rendering it almost impossible to quickly understand the violation.
Consequently, system designers might write down the values in a table-like representation (\autoref{fig:cur-workflow-cex}b).
Only then they can start to relate formula, system, and the counterexample executions with each other in order to identify and understand the violation of the specification.

\subsection{Challenges}
\label{sec:challenges:challenges}
This whole process quickly becomes cumbersome and poses multiple challenges.
First of all, hyperproperties can express 
arbitrarily complex relations across traces and time, making it hard to recognize the patterns in the executions that violate such a hyperproperty.
Further, analysts need to identify which subformulas were relevant (i.e., violated) and which parts can be ignored.
Any visual support for scanning the present executions and identifying the relevant elements will be highly beneficial.
After identifying the violation, it is still necessary to reason about \emph{why} the violation could occur, thus, why one execution reached a particular state.
For this, the executions must be considered in the context of the provided system.
Further, both formula and system can grow quickly in size.
On the one hand, this makes it increasingly challenging to sketch the components adequately and recognize specific characteristics.
On the other hand, the likelihood of faulty formula specifications or system definitions increases as well, leading to the need to identify these issues and to correct them.
Thus, editing facilities for formula and system are of interest.
With our tool, we aim to significantly improve these analysis workflows and help the system designers in their development process.

%

\section{Background \& Related Work}

We first provide formal details of the model checking problem of hyperproperties.
Secondly, we elaborate on the importance of visualization methods to better understand abstract models or processes by giving an overview of related work.
Finally, we discuss existing work for editing formula and graph representations.

\subsection{Model Checking of Hyperproperties}

Model checking~\cite{10.5555/332656} answers the following question: Given a system description $M$ and a specification $\varphi$, formally describing the desired property, does $M$ satisfy $\varphi$. More specifically in the context of hyperproperties, we require that the set of executions of $M$ satisfies the hyperproperty. For the interested reader, we will define these concepts formally in the following.

The system description $M$ is typically provided as a finite Moore state machine, formally defined as a tuple $(S,s_0,I,O,\tau,l)$ with: $S$: a finite set of states; $s_0$: the initial state; $I$: the input alphabet; $O$: the output alphabet; $\tau: S \times I \rightarrow S$: a transition function; and $l: S \rightarrow O$: an output labeling.
Figure~\ref{fig:info-leak}, for example, depicts a finite Moore state machine with three states. The input alphabet contains variables $i$ and $s$ and the output alphabet contains the variables $o_1$ and $o_2$. Edges of the state machine (arrows) are labeled with the input and states (circles) are labeled with the system's output.
An execution (trace) of a model~$M$ is an infinite sequence of sets of atomic propositions $\mathit{AP}$ through the state machine, where $\mathit{AP} = I \cup O$.
An example trace of the model in Figure~\ref{fig:info-leak} is $\{i,s\}(\{o_2\})^\omega$. 
In the first position of the trace (corresponding to the initial state and first input), there is no output but the input $i$ and $s$.
Defined by the transition function, we proceed from the initial state to state $2$, where we reside indefinitely by outputting $o_2$ without receiving a further input. 
The notation $(x)^\omega$ denotes that $x$ is repeated infinitely often.
Formally, the set of all traces for a set of atomic propositions is thus $(2^\mathit{AP})^\omega$, i.e., the set of above mentioned infinite sequences over atomic propositions.
The set of traces of a system model $M$, denoted by $\mathit{Traces}(M)$, is a subset of $(2^\mathit{AP})^\omega$.

Formally, a hyperproperty $H$ is a set of sets of traces; meaning it defines all trace sets that comply to the hyperproperty.
If the traces of a system model $M$ are no element of the hyperproperty, i.e., if $\mathit{Traces}(M) \not \in H$ then the system does not satisfy the hyperproperty.
In this case, a counter example is provided by the model checker, i.e., a set of system traces that violates the hyperproperty.

The desired behavior of the system is provided in a formal specification language such as HyperLTL, a temporal logic for hyperproperties.
In HyperLTL, variables are interpreted as atomic propositions which can be connected with either Boolean operators (e.g., \emph{equivalence} $\leftrightarrow$, \emph{implies} $\rightarrow$, \emph{or} $\lor$) or temporal operators.
The most prominent temporal operators are \emph{globally} ($\G \varphi$, where $\varphi$ must be true at all times) and \emph{eventually} ($\F \varphi$, meaning that $\varphi$ will hold at some point in time); further operators include \emph{until} ($\U$), \emph{release} ($\R$), and \emph{next} ($\X$).
As an example, consider again the \hyperltl formula from \autoref{sec:challenges:example}: $\forall\pi ~ \forall\pi' ~ \G \left(i_\pi \leftrightarrow i_{\pi'}\right) \rightarrow \G \left(o_\pi \leftrightarrow o_{\pi'}\right)$.
HyperLTL formulas start with a quantifier prefix introducing universally ($\forall$) or existentially ($\exists$) quantified trace variables ($\pi$ and $\pi'$)
followed by a formula $\psi$ in the body (here $\G \left(i_\pi \leftrightarrow i_{\pi'}\right) \rightarrow \G \left(o_\pi \leftrightarrow o_{\pi'}\right)$).
Within this formula, the variables are indexed with trace variables to indicate to which trace quantifier they refer to (e.g., $i_\pi$).
For a formal definition of the semantics and more examples of hyperlogics, we refer the interested reader to~\cite{Clarkson2014TemporalLogicsHyperproperties,Coenen2019HierarchyHyperlogics}.

With \hypervis, we visualize hyperproperty counterexamples returned by a hyperproperty model checker~\cite{Finkbeiner2015AlgorithmsModelChecking,DBLP:conf/cav/CoenenFST19,DBLP:conf/cav/FinkbeinerHT18}.
We use \mchyper~\cite{Finkbeiner2015AlgorithmsModelChecking}, which builds on \abc~\cite{Brayton2010ABCAcademicIndustrial}.
\mchyper takes as inputs a hardware circuit, specified in the \aiger format~\cite{Biere2007AIGERInverterGraph, Biere2011AIGER1.9}, and a \hyperltl formula.
\mchyper solves the model checking problem by computing the self-composition~\cite{Barthe2011SecureInformationFlow} of the system.
If the system violates the \hyperltl formula, \mchyper returns a counterexample.
This counterexample is a set of traces through the original system that together violate the \hyperltl formula.
Depending on the type of violation, this counterexample can then be used to debug the circuit or refine the specification iteratively.

\subsection{Visualization and Explication of Formal Methods}
In recent years, research started more intensively to investigate how complex and abstract algorithms and models can be visualized and interactively explored, and, thus, be made more transparent.
Most prominently, this includes work within the area of \emph{explainable artificial intelligence} (XAI)~\cite{Gunnning2016Explainable, Spinner2020explAIner, Strobelt2019Seq2seq}, but can also be extended to related fields such as formal methods~\cite{Flemisch2020Towards}.
For example, proof attempts have been visualized by \tool{SatVis}~\cite{Gleiss2019InteractiveVisualizationSaturation} and an improved version of the Z3 Axiom Profiler~\cite{Rothenberger2016IntegrationAnalysisAlternative}.
They visually represent attempts from the \tool{Vampexcire} theorem prover and Z3 SMT solver, respectively, in order to support users and developers of the tools in understanding the results.

\paragraph{Textual Explications}
One instance of textual explications are automatically generated facts based on the underlying data~\cite{Srinivasan2019AugmentingVisualizationsInteractive}.
Typically, machine learning algorithms extract facts which are then verbalized using natural language generation (NLG)~\cite{Spreafico2020NeuralDataDriven,Obeid2020Chart,Liu2020AutoCaption}.
These facts can then aid interpreting a visualization by verifying the viewer's thoughts and pointing at potentially overlooked aspects~\cite{Srinivasan2019AugmentingVisualizationsInteractive}.
The generated facts can be shown as a single caption for a chart~\cite{Shi2021Calliope} or be provided as a collection of statements next to the visualization~\cite{Cui2019DataSite}.
Applications in the areas of student-teacher communication~\cite{MartinezMaldonado2020FromData}, XAI~\cite{Hohman2019TeleGam,Sevastjanova2018Going}, and supporting safe handovers in cyber-physical systems~\cite{Wiehr2020SafeHandoverMixed} further indicate their practical benefits for interpreting visualizations.

\paragraph{Visualizing Counterexamples}
Visually representing counterexamples for trace properties, e.g., for \tool{LTL}, is a known challenge for which various approaches have already been proposed.
Techniques, such as state diagrams~\cite{Jeong2010VISAnalyzerVisual, Loer2006IntegratedFrameworkAnalysis, Aljazzar2008DebuggingDependabilityModels, Goldsby2006VisualizationFrameworkModeling}, sequence diagrams~\cite{Loer2006IntegratedFrameworkAnalysis, Campetelli2011UserfriendlyModel,  Larsen201820YearsUPPAAL}, and variable tables~\cite{Jeong2010VISAnalyzerVisual, Loer2006IntegratedFrameworkAnalysis, Pakonen2018CounterexampleVisualizationExplanation}, convert the counterexample and the system model to more readable formats.
The model view~\cite{Goldsby2006VisualizationFrameworkModeling, Larsen201820YearsUPPAAL, Campetelli2011UserfriendlyModel} takes a different route by mimicking the counterexample and providing a step-wise navigation.
Further, visualization approaches for such counterexamples with single executions have been considered for various domains and applications~\cite{Bolton2010UsingTaskAnalytic, Jee2010FBDVerifierInteractiveVisual,  Patil2015CounterexampleguidedSimulation, Bochot2010PathsPropertyViolation}.
Additionally, the established model checker \tool{UPPAAL}~\cite{Larsen201820YearsUPPAAL} visualizes timed automata for real-time systems, allowing for interacting with simulations of the system.

Approaches for supporting the analysis of counterexamples include minimizing~\cite{Lahtinen2012ModelCheckingMethodology} and explaining counterexamples~\cite{Beer2009ExplainingCounterexamplesUsing}, as well as investigating several system executions simultaneously~\cite{Groce2003WhatWentWrong, Schuppan2005ShortestCounterexamplesSymbolic, Bochot2010PathsPropertyViolation}.
Multiple works explore how individual counterexamples can be visualized and explained, e.g., for function block diagrams~\cite{Jee2010FBDVerifierInteractiveVisual, Pakonen2018CounterexampleVisualizationExplanation}, with the newest version of \tool{MODCHK}~\cite{Pakonen2018CounterexampleVisualizationExplanation} being highly related to \hypervis.
\tool{MODCHK} provides a causality analysis~\cite{Beer2009ExplainingCounterexamplesUsing} which delivers an over-approximation of a set of causes.
In contrast, \hypervis produces minimal explanations using a more efficient explanation algorithm.
Further approaches to identifying the causes of a \emph{trace property} violation~\cite{Groce2003WhatWentWrong} have been made, for instance, the \textsc{explain}~\cite{Groce2004UnderstandingCounterexamplesExplain} tool, which has been incorporated into multiple model checkers~\cite{Clarke2004ToolCheckingANSI, Chaki2004ModularVerificationSoftware, Chaki2004ExplainingAbstractCounterexamples}.

\paragraph{Visualizing Parallel Executions}
Research on distributed systems has examined how to visualize multiple, parallel executions of a system.
Two examples are \tool{Oddity}~\cite{Michael2019TeachingRigorousDistributed, Woos2019StepDebuggerDistributed} and \tool{ShiVis}~\cite{Abrahamson2014SheddingLightDistributed, Beschastnikh2020VisualizingDistributedSystem}.
Oddity consists of an interactive visual debugger and is part of the \tool{DSLabs} framework, which also includes a model checker for distributed systems.
\tool{ShiVis} is a web tool that uses space-time diagrams to visualize the execution of a distributed system.
Particularly these diagrams highlight the communication between components and the partial ordering between events that happen across components (the happens-before relationship).
However, while related investigations have been conducted, the specific case of visualizing model checking results of hyperproperties was not considered.

\subsection{Editing Formulas and Systems}

The modeling of systems that fulfill certain specifications is an iterative process of checking, correcting, and refining both the specifications and the system models. 
Therefore, editing the specification, i.e., formula or system, are essential parts of the workflow.
In the following, we present existing work providing techniques for efficiently editing them.

\paragraph{Formula Editing}
Established online tools like Wolfram Alpha \cite{Wolfram-SMC} feature advanced text editors that facilitate writing mathematical notations.
Specifically for formula editing, most interfaces provide a real-time preview of the formula, translated from the markup language used for writing the mathematical expressions.
The most common markup languages for mathematical input are \LaTeX{} \cite{Gratzer-MathIntoLatex, WEB:SeoulOh2012MathQuill} as well as OpenMath and MathML \cite{kohlhase2012semantics, MathMLV3-2014}.
Via markup alternatives or special characters keyboards, WYSIWYG approaches can allow users without knowledge of the markup language to still write the desired mathematical expressions~\cite{LaoTebar2016ProposalFC, Pollanen2014TowardsAU}.
Such visual interfaces can also support focusing on specific formula parts by collapsing selected subformulas~\cite{cervone2015towards, Kohlhase08adaptationof}.
Finally, more experimental interfaces are starting to provide handwriting and speech recognition capabilities~\cite{WigmHuntPflu2009dp, LaoTebar2016ProposalFC, DeepNeuralNetHandwritingRecognition}. 

\paragraph{System Editing}
System models are usually edited in a hardware description language like \tool{Verilog}~\cite{IEEE-Verilog} within an integrated development environment.
As an alternative to these textual representations, the systems can also be modeled through finite state machines \cite{HDL-FSM, FSM-to-VHDL, New-FSM-to-VHDL}. 
These models can then be visually edited, e.g., by adding, relocating and removing nodes or edges from the node-link diagram \cite{Frisch2012Node-link-editing, Romat2021ExpressiveAuthoringNode}.
Lightweight versions of such editing are already provided within commercial tools for general diagram editing, such as Stateflow~\cite{Simulink}.

%

\section{HyperVis: Visualizing Model Checking Results}
Based on the identified challenges of analyzing model checking results (\autoref{sec:challenges:challenges}), we iteratively developed \hypervis.
In this section, we will first recap the main components of the considered counterexamples (\autoref{sec:hypervis:components}) and outline our set design goals (\autoref{sec:hypervis:designgoals}).
Then, as the main part, we will present our visualization design (\autoref{sec:hypervis:design}), including its interaction concepts.
This is followed by the description of the considered editing and debugging facilities (\autoref{sec:hypervis:tool-editing}).
Finally, we provide further insights into the design process as well as the actual implementation (\autoref{sec:hypervis:implementation}).
The tool is provided online (\href{https://imld.de/hypervis}{\color{href-blue}imld.de/hypervis}).

\subsection{Components of Counterexamples}
\label{sec:hypervis:components}

Strictly speaking, a counterexample to a hyperproperty is only the set of executions that are returned by the model checking tool.
However, for the remainder of this work, we depict a counterexample to comprise the formula and system provided by the analyst as well.
Thus, it consists of three main components: the \emph{system}, the \emph{formula}, as well as the specific \emph{executions}.
In addition, we introduce \emph{explanations} as a fourth component, indicating and explicating relevant bits of the violation.

The system describes the hardware circuit as a transition system with states providing the \emph{outputs} and transitions implementing state changes based on \emph{inputs}.
Due to the system being a hardware circuit, states are internally represented by \emph{latches}, i.e., sub-circuits that can preserve information.
Together, all available variables, i.e, outputs, inputs, and latches, are the \emph{atomic propositions}.
The formula can be represented as a syntax tree over propositional and temporal \emph{operators} where leaves are selected atomic propositions on a specified \emph{trace} (or \emph{execution}).
Here, a formula typically describes relations on pairs of executions, i.e., two instances of the system.
Each execution is representing values of atomic propositions for every \emph{time step}.
Notably, these executions can be infinite and contain a \emph{lasso} (or loop), which marks subsequent time steps that are repeated infinitely.

Through a causal analysis of the counterexample, we are able to identify which atomic propositions in the formula contributed to its overall violation and are, therefore, \emph{relevant} for the counterexample.
For the explanations, we extract textual explications of the most top level \emph{subformulas} with temporal operators.
Depending on the actual top level operator, these subformulas can either be satisfied or violated (e.g., in case of an `implies' $\rightarrow$ operator, the premise has to be satisfied while the conclusion is then violated).
For an analyst, all mentioned components and elements are relevant for understanding the counterexample in general, reasoning about why it can occur, and identifying possible corrections to either the formula or system.

\begin{figure*}[!t]
    \centering
    \includegraphics[width=\textwidth]{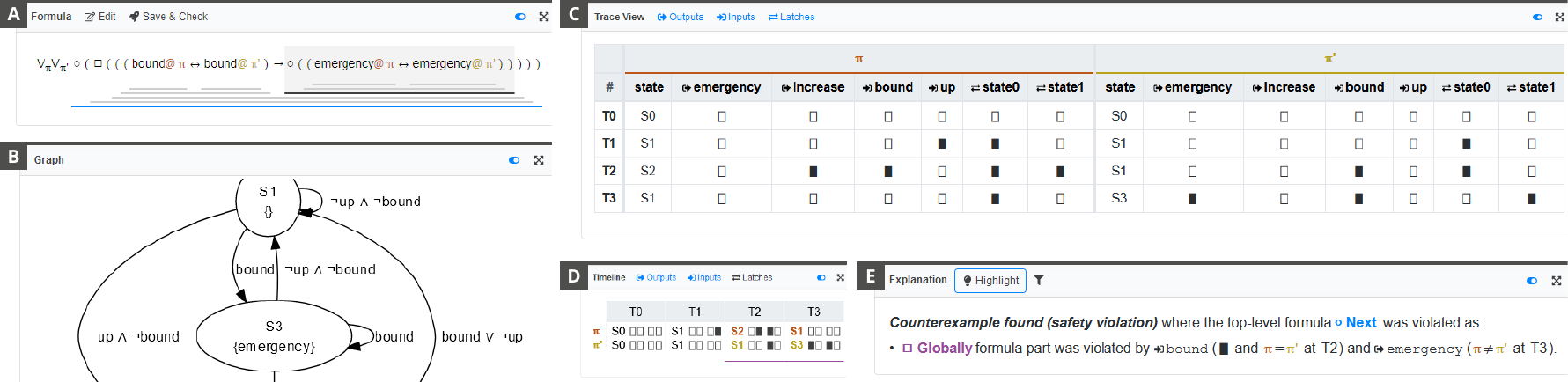}
    \vspace{-17pt}
    \caption{Views of \hypervis: (a) \formulaView, here with hover of a subformula; (b) \graphView showing the system as a Moore transition system, here zoomed in; (c) \traceView providing for both executions {\color{trace0}$\pi$} and {\color{trace1}$\pi'$} the values of all atomic propositions across all time steps; (d) \timelineView showing the executions in a compact format; and (e) \explanationView with textual statements on the counterexample, here with one relevant subformula.}
    \label{fig:views}
    \vspace{-10pt}
\end{figure*}

\subsection{Design Goals}
\label{sec:hypervis:designgoals}

When starting the design process, we identified multiple design goals that a tool for visually analyzing hyperproperty counterexamples should fulfill.
The goals DG1--DG5 describe desired visualization aspects, while DG6--DG7 outline more general tool characteristics.

\begin{description}
\itemsep0em 
    \item[DG1: Build Upon Familiar Presentations.] As illustrated in \autoref{fig:cur-workflow-system+formula} and \autoref{fig:cur-workflow-cex}, analysts often sketch the system or list the executions in a certain way.
    We aim to foster an intuitive understanding of the views by building upon these typical representations, but extending them with more effective encoding strategies.
    \item[DG2: Support Recognizing Trace Relations.] In many cases, a specific combination of absent or present atomic propositions must be identified and compared across the executions.
    We aim to simplify such pattern recognition within the executions.
    \item[DG3: Relating Components.] A major challenge for analysts is relating the different components to each other, e.g., mapping back atomic propositions in the execution to corresponding subformulas or to taken transitions in the system.
    Thus, the tool should support the analyst in mentally linking elements across views.
    \item[DG4: Provide Guidance for Identifying Violations.] Counterexam\-ples can quickly become overwhelming, with a multitude of variables or time steps being involved.
    Our goal is to support analysts in identifying the relevant elements that led to the violation and, thus, in understanding the model checking result.
    \item[DG5: Enable Editing of Formula and System.] Due to their complexity, formulas and systems can easily contain small but hard to recognize bugs leading to a counterexample.
    For this, the tool should provide integrated functionalities for fixing such issues.
    \item[DG6: Provide a Holistic Interface.] Model checking of hyperproperties is a multi-step process; from providing the input, to analyzing the counterexample, to iterating the specification or system.
    Thus, a tool should consolidate these steps within one interface.
    \item[DG7: Avoid Setup Efforts.] The tools used for model checking are often command-line based and implemented with different dependencies.
    We aim at avoiding the setup effort for the analyst and providing a unifying ready-to-use tool.
\end{description}
In the following, we detail how we addressed these design goals within the visualization design of \hypervis' views.

\subsection{Visualization Design}
\label{sec:hypervis:design}

Guided by the described design goals, we developed \hypervis and its general interface including five visualization views.
The focus in this section lies on how we visualize the counterexample components specifically as well as efficiently guide the analysis.

\subsubsection{Visualizing a Counterexample: Provided Views}
For \hypervis, we developed five different views; the \formulaView, \graphView, \traceView, \timelineView, and \explanationView.
In the following, their design is detailed.

\paragraph{\FormulaView}
The \hyperltl formula provided by the user is transformed into a representation using the actual logical and temporal operator symbols (DG1).
Internally, the formula is in a hierarchical structure; this structure is indicated with bars below the formula string.
The bars allow emphasizing the different subformula levels, with the uppermost bars representing the atomic propositions and the lowermost bar (marked in blue) the entire formula.
Hovering over the bars emphasizes the corresponding subformula (\autoref{fig:views}a), simplifying recognizing the formula structure and corresponding bracket.
The stated atomic propositions always relate to one specific trace.
To simplify distinguishing which proposition corresponds to which trace, we introduced fixed colors for the traces and added labels to the proposition, i.e., either {\color{trace0}$@\pi$} or {\color{trace1}$@\pi'$}.
These trace colors are re-used in all views.

\paragraph{\GraphView}
The system is visualized as a Moore transition system, i.e., a graph with the states as nodes and transitions as edges (\autoref{fig:views}b).
Following their convention, the set of present outputs on a given state is printed into the node label, e.g., $\{emergency\}$ in state S3.
If an output is absent, its value is \emph{false}.
Further, we show symbolic transitions, i.e., edges can be labeled with formulas expressing specific input combinations, such as logical conjunction (e.g., $up \land \neg bound$).
The graph can be freely zoomed and panned.

\paragraph{\TraceView}
As previously described, analysts typically transform the textual output of the model checker into a table-like format, thus creating an overview of all atomic propositions and their values on the traces across time steps.
Our \traceView builds upon that (DG1) and prints the atomic propositions per trace as columns and the time steps as rows (\autoref{fig:views}c).
The values themselves are binary, thus are either $\true$ when a variable was present or $\false$ when it was absent.
We propose to replace the common notation of the values as $1$ and $0$ with a graphical representation: a filled rectangle \inlineTrue represents present variables and a hollow rectangle \inlineFalse absent variables.
This representation simplifies recognizing patterns of occurring values in and across traces (DG2).

Small icons before the proposition name indicate its type, i.e., either \inlineOutput output, \inlineInput input, or \inlineLatch latch.
The propositions are sorted first by type and then alphabetically.
Controls in the view head allow for hiding an proposition type.
In addition to the atomic propositions, we also show a numbered state indicator (e.g., S3) in the first column of each trace.
These state indicators are abstractions of the latches, which together encode the current state.
The time steps are labeled as T0, T1, and so forth.
Further, if a lasso (see \autoref{sec:hypervis:components}) is present in the counterexample, it is indicated with gray borders at the respective time steps.

\begin{figure*}
    \centering
    \includegraphics[width=\textwidth]{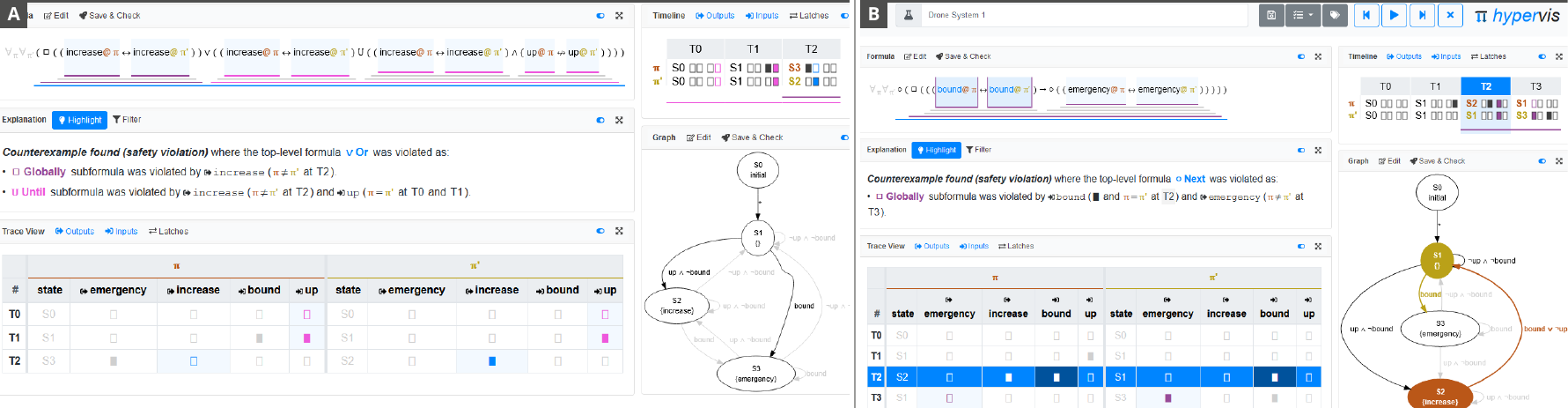}
    \vspace{-17pt}
    \caption{The interactive analysis of counterexamples is at the core of \hypervis: (a) activated explanation highlighting points analysts to relevant elements; (b) stepping through time steps fosters the understanding of the executions' behavior (controlled by the buttons in the top right corner).}
    \label{fig:analysis}
    \vspace{-10pt}
\end{figure*}

\paragraph{\TimelineView}
So far, the view design was influenced by common ways to write down the counterexample.
However, the \timelineView is a new visualization that aims to provide a more compact representation of the executions (\autoref{fig:views}d).
Similarly to the \traceView, it shows the specific values of the atomic propositions, but with the time mapped horizontally.
By omitting the atomic proposition labels, the rectangles indicating present or absent variables are placed next to each other.
This allows for a further improved pattern recognition, either across traces or across time steps (DG2).
For example, considering a set of four variables, it is easily possible to observe differences or similarities across instances: \inlineFalseNSp\,\inlineTrueNSp\,\inlineTrueNSp\,\inlineFalse and \inlineFalseNSp\,\inlineFalseNSp\,\inlineTrueNSp\,\inlineFalse differ only in the second variable.
The label of the represented proposition can be accessed by hovering over the rectangle; their order is equal to the order in the \traceView.

In an earlier iteration, the view was intended to emphasize diverging behaviors of the executions for one variable, e.g., showing when they were in different states or read different values for an atomic proposition.
We opted to develop the view further into its more compact format while also showing the atomic propositions.
To still indicate diverging executions, the state indicator (e.g., S0) is colored black if both executions are in the same state and colored according to the trace color when they diverge (e.g., {\color{trace0}S2} and {\color{trace1}S1}).
Finally, in the case of a present lasso, an indicator at the time steps is provided.

\paragraph{\ExplanationView}
The \explanationView shows a verbal summary of the counterexample alongside statements on the most top-level subformulas relevant to the found violation (\autoref{fig:views}e).
The basis of this is an automated causal analysis of the counterexample, which extracts a minimal set of subformulas that contributed to the overall violation at one or more time steps.
With this information, we can relate the subformulas to specific values at specific time steps and derive a textual statement.
The statements' structure is always the same: first, the temporal operator of the subformula is stated, followed by a list of involved atomic propositions.
For each proposition, it is indicated at which time step it became relevant, how the values relate to each other across traces
and whether the values were always $\true$ or $\false$.
This information is provided as inline or word-sized representations~\cite{Beck2017WordSizedGraphics, Goffin2017ExploratoryStudyWord}, seamlessly integrating into the textual description.
Further, each statement is assigned a unique color, allowing for indicating it in other views (DG3).
For example, as visible in \autoref{fig:views}d, the \timelineView shows bars at the bottom, hinting at which time steps a subformula was relevant.

A complete statement is shown in \autoref{fig:views}e.
For $bound$ the value was $\true$ (\inlineTrue) and equivalent on both traces ({\color{trace0}$\pi$}$=${\color{trace1}$\pi'$}) at T2, while $emergency$ was unequal on both traces ({\color{trace0}$\pi$}$\neq${\color{trace1}$\pi'$}) at T3.
These statements can help analysts to quickly grasp the essential aspects of the violation and locate the time steps and atomic propositions of interest.
Currently, the provided explanations for the atomic propositions can indicate found equivalences of traces across $n$ time steps as well as consistent values across time steps (DG2).
However, as \hyperltl formulas can describe arbitrary relations of atomic propositions across traces, not all relevant patterns are currently recognized and expressed.
Similarly, we only provide statements for subformulas being present at the top two levels; thus, more nested formulas might not be verbalized adequately.

\paragraph{Interface Arrangement}
By default, the views are arranged in a simple 2-column grid, with formula, explanation, and \traceView being placed in a wider column on the left, and timeline and \graphView on the right (\autoref{fig:teaser}).
However, as the space requirement of the views can heavily vary between counterexamples, the interface also supports arranging the views differently.
For example, if a formula is becoming rather long, it is placed in full width on top.
Similarly, if counterexamples involve many time steps, \traceView and \timelineView are devoted more space.
In general, the goal is to provide all views within the initial viewport and avoid scrolling as much as possible.
Finally, analysts can also manually collapse or maximize views.

\subsubsection{Analyzing a Counterexample: Interactive Guidance}

For analyzing the counterexample, we provide further interactive mechanisms fostering the comprehension of the counterexample's specifics.
These mechanisms include an explicit highlighting of relevant elements, linked highlighting across views, as well as a debugger-like stepping through the counterexample.
These dynamic functionalities of \hypervis are also shown and explained in the accompanying video.

\paragraph{Highlighting Relevant Elements}
As stated in the context of the \explanationView, we are identifying the subformulas that contributed to the overall violation at specific time steps.
This knowledge is not only used for the explanation statements but also to indicate the relevant elements across all views (DG4).
To activate this indication, the \explanationView features a `Highlight' toggle button (\autoref{fig:views}e).
Upon activation, as in \autoref{fig:analysis}a and \autoref{fig:teaser}, the non-relevant subformulas in the \formulaView are grayed out, as are the non-taken states and transitions of the executions in the \graphView.
Similarly, non-relevant values in the \traceView and \timelineView are shown less opaque while relevant values are emphasized.
Further, a filter button next to the highlight button allows for removing non-relevant elements from the views.

We also relate the relevant elements to the provided statements in the \explanationView (DG3).
Specifically, we identify which atomic proposition is part of which statement, i.e., in which subformula it occurs.
Further, we propose to use the statements' assigned color for highlighting: the rectangle representing binary values are colored accordingly, as are the bars indicating the atomic proposition.

\paragraph{Linked Highlighting}
In general, all views react to hovering over displayed elements, e.g., subformulas, states, or time steps.
Hovering also results in a linked highlighting across views~\cite{Buja1991InteractiveDataVisualization, Roberts2007STAR_CMV}, i.e., the corresponding elements in other views are also highlighted (DG3).
Only in a few cases, elements are shown in exactly the same way in other views.
For example, subformulas in the \formulaView may occur again the \explanationView.
As in most cases elements appear slightly differently, e.g, \formulaView and \timelineView show atomic propositions differently, the correspondence is not immediately apparent and is then indicated through the linked highlighting (DG3).

Specifically, hovering over a trace indicator (i.e., {\color{trace0}$\pi$} or {\color{trace1}$\pi'$}) in either view highlights the execution in the \graphView, i.e., all taken transitions and states are colored in the trace color.
Vice versa, hovering a state or transition highlights instances in the executions where this state and the inputs of the transition were present.
Hovering over a time step highlights the corresponding row or column in trace and \timelineView, the relevant subformulas at this time step, and the executions' current states and transitions taken next.
The atomic propositions in formula and \explanationView allow for highlighting the corresponding labels in the \traceView and, if applicable, the specific values that were relevant at certain time steps in both trace and \timelineView (DG4).

\paragraph{Stepping through a Counterexample}
It is important to understand the sequence of events that lead to the violation.
Therefore, we enable stepping through the counterexample in a debugger-like fashion (\autoref{fig:analysis}b).
Through control buttons provided in the interface header, the analyst can move forward and backward.
For the current step, a stronger visual highlight is used  (\autoref{fig:analysis}b), with the time step colored in blue and relevant subformulas further emphasized.
For the \graphView, we color the states and transitions in the respective trace colors; if they share the same state or transition, blue is used.
The same effect can also be achieved by selecting a time step in the \traceView or \timelineView.
Further, when stepping through, the highlight is permanent and can be used in combination with the highlighting of relevant elements as well as the linked highlighting triggered on hover.

\subsection{Tool Functionalities \& Editing Facilities}
\label{sec:hypervis:tool-editing}

Following DG6, we provide one unified interface that allows for performing model checking, analyzing the counterexamples, and iterating specification and system within it.
In the following, we describe the tool functionalities of \hypervis as well as its editing facilities.

\paragraph{Tool Functionalities}
We extended \hypervis with tool functionalities allowing for using it in a productive way for many model checking projects simultaneously.
Among others, this includes functionalities for (re-)loading projects, re-running model checks, or managing different versions of them.
The project manager provides access to all model checking projects, i.e., loaded systems and specifications checked, in a sidebar widget.
The different projects can have multiple versions (here, marked with the latest modification timestamp), helping analysts to quickly jump to older iterations.
For each version, multiple checks can be created, i.e., multiple hyperproperties that a system should fulfill.
The versions of the projects can be manually created, but are also automatically introduced when editing a formula or system.
In case of faulty edits that result in an error thrown by the model checker, a rollback to the previous version is offered.
Each version of the project can be freely tagged and each checked hyperproperty can be named.

\begin{figure}[!t]
    \centering
    \includegraphics[width=\columnwidth]{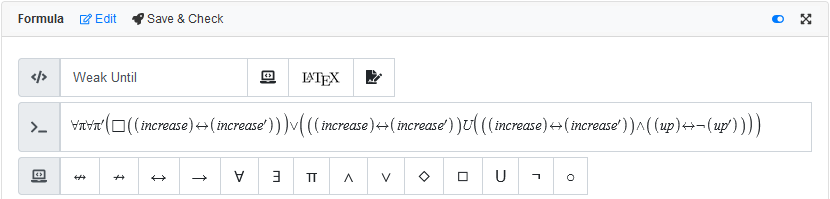}
    \vspace{-17pt}
    \caption{Within the \formulaView, the formula editor can be toggled, providing a WYSIWYG inline \LaTeX{} editor.}
    \label{fig:formula-editor}
    \vspace{-10pt}
\end{figure}

\paragraph{Editing of the Specification}
\hypervis provides a formula editor that uses \LaTeX{} markup to edit the hyperproperty specifications (\autoref{fig:formula-editor}).
Specifically, the analyst can directly type \LaTeX{} markup, which is automatically rendered inline.
By using \LaTeX{}, copying and pasting formulas edited in common tools outside of \hypervis becomes directly possible.
Similarly, formulas could also be sketched by hand via pen input. However, this is currently only supported by external tools like \tool{MyScript}~\cite{WEB:MyScript2017}.
Our editor also features an extended keyboard, providing access to the common logical and temporal operators.
As mentioned before, \hypervis supports the editing of multiple separate formulas for a single system, potentially allowing to split up complex hyperproperties or to test very different specifications.

\paragraph{Editing of the System}
For editing the system model, changes can be made either through visual editing or by changing the original text input.
The visual editing could involve providing a special mode allowing for, e.g., drawing edges, rerouting existing ones, or creating nodes.
At the same time, some users might still prefer to directly textually edit the originally provided \tool{Verilog} definition.
However, as providing such editing supporting is not straightforward, \hypervis currently only features a mock-up editing of the system.
Specifically, the challenge comes with the representation of the system as a hardware circuit in the \aiger format~\cite{Biere2011AIGER1.9}.
These \aiger files are automatically generated from definitions implemented in \tool{Verilog} and are hardly readable by humans.
For displaying the \graphView, a DOT~\cite{Koutsofios1992DrawingGraphsDot} representation is generated from \aiger, however, the transformation comes with information loss and is therefore not invertible.
To sidestep this, an intermediate format for automatons could be used, which can be transformed from or to \aiger without information loss and also more easily changed in programmatic way.
For now, we allow the editing of the DOT notation, which updates the shown \graphView to illustrate the intended functionality but cannot trigger an updated model check.

\begin{figure*}
    \centering
    \includegraphics[width=\textwidth]{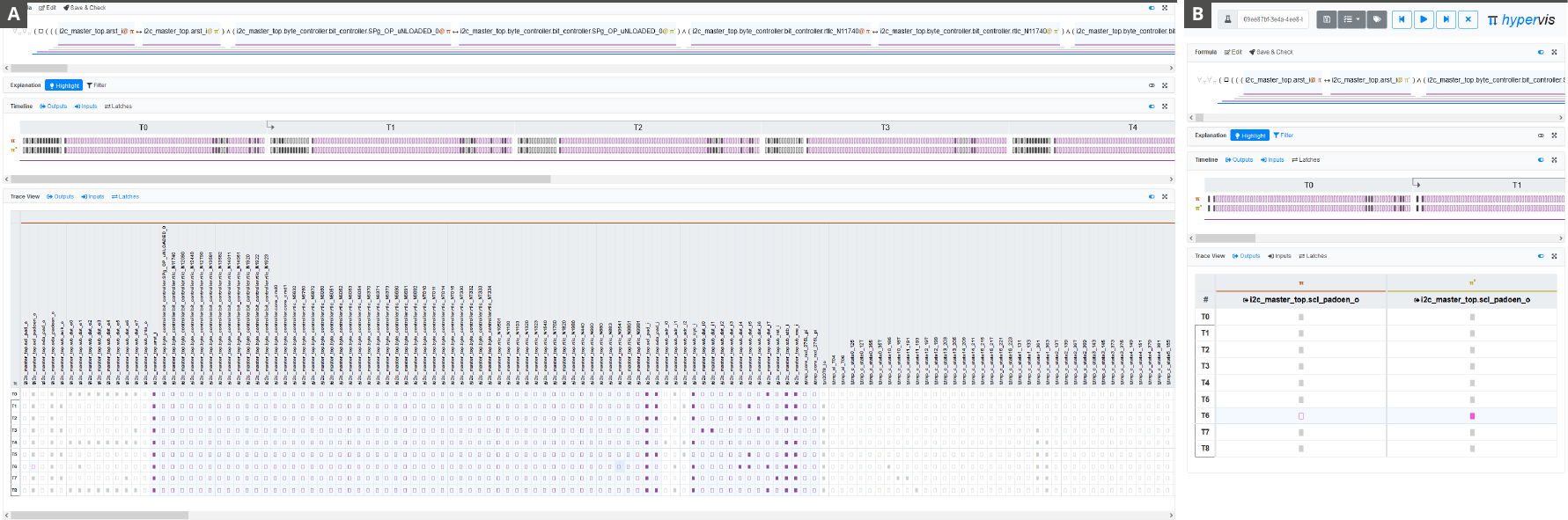}
    \vspace{-17pt}
    \caption{
    (a) Excerpt of the interface for the I2C case study with activated explanation highlighting, where the \traceView can only show a fraction of the 680 columns (see scroll bar); (b) with the explanation filtering activated as well as latches and inputs disabled, the \traceView can be limited to the one relevant output variable. In both views, the explanation view has been collapsed; the full view is provided in the supplement.}
    \label{fig:i2c}
    \vspace{-10pt}
\end{figure*}

\subsection{Design Process \& Implementation}
\label{sec:hypervis:implementation}

In order to develop \hypervis, we followed an iterative design process within an interdisciplinary team.
This team consists of formal methods researchers on the one hand and HCI as well as visualization researchers on the other hand.
While not end-users, the first group are domain experts for model checking of hyperproperties, knowing the challenges and main goals.
In the following, we detail the design phases as well as the current implementation of \hypervis.

\subsubsection{Design Phases}
The first phase involved introducing the visualization researchers to the domain of model checking and hyperproperties in order to establish a common understanding of the current processes and present challenges.
Afterward, we jointly developed a first click prototype illustrating a possible interface visualizing the found counterexamples.
Then, the first implementations of the visualizations were realized in a web prototype alongside a parser consuming the output file of the model checking generated by \mchyper.
Within this process, it became apparent that plainly representing the counterexample will be insufficient and that it will be essential to become able to extract the relevant bits of the violation and presenting it to the analyst.
At this stage, a first version of the causal analysis algorithm was developed alongside early highlighting mechanisms.
This enabled testing various case studies, and thus incorporating further improvements into the visualization design.

On the tool side, we started to develop approaches for editing the formula and system as well as the general structure of the tool interface, e.g., providing access to the project list, their versions, or loading new ones.
With these tool aspects implemented, we ran a first feedback session with 3 participants and collected comments on the interface.
This feedback allowed us to iterate, e.g., the menu structures, button icons and labels, or features of the inline formula editor.
The result of the overall design process is the current version of \hypervis.

\subsubsection{Implementation as Web Tool}

\hypervis is implemented as a web-based tool, featuring a \tool{Node.js}-based~\cite{WEB:OpenJSFoundation2009Node.js} backend and JavaScript-based frontend.
In the frontend, the views are implemented with plain HTML or SVG.
Except for the \explanationView, the rendering of all views is controlled by \tool{D3.js}~\cite{Bostock2011D3DataDriven}.
To support the linked highlighting, custom events were introduced that are sent and consumed by the views.
For the formula editing, we incorporate the \tool{MathQuill} library~\cite{WEB:SeoulOh2012MathQuill}.
For translating the formula and producing the polish notation that \mchyper requires, we use \tool{Spot}.
Graph editing is not fully functionally implemented yet.
To illustrate the general possibility, we provide an embedded \tool{CodeMirror}~\cite{WEB:Haverbeke2011CodeMirror} editor to change the DOT representation of the system.

The backend is responsible for managing the model checking pipeline.
Based on the analyst's inputs, it calls the \mchyper Python tool before handing over the found counterexample to our own Python script extracting the relevant subformulas.
It computes a minimal set of variable and time step pairs, which cause the violation.
In addition, this script also writes all required information into a JSON file.
In parallel, a separate script is used to generate the DOT representation based on the \aiger file.
As for larger systems this generation might not terminate in a reasonable time, for some counterexamples the graph is not available.
After parsing these generated outputs in the \tool{Node.JS} server, the data is provided to the frontend.
For each project, the results are stored in a folder structure, allowing to quickly reload the counterexamples later on and implement the versioning concept.
The communication between the frontend and backend is based on HTTP requests.

The \hypervis tool is hosted online but can also be locally run, either as a Docker container or by fully installing it and all its dependencies.
In general, we envision the usage as an online tool as the primary usage style, which then also allows for avoiding setup efforts (DG7).
\section{Validating \hypervis}
In this section, we validate \hypervis by discussing multiple case studies and reporting on user feedback sessions.
Both illustrate that \hypervis indeed advances the state-of-the-art significantly by helping to quickly identify the violations in counterexamples of hyperproperties.

\subsection{Case Studies}
Here, we detail two selected case studies:
In the first one, we visualize the results of model checking information-flow properties on an open source implementation of the I2C bus protocol.
In the second case study, we take a look at one of the core building blocks of such bus implementations: mutual exclusion protocols.

\paragraph{CS1: I2C Bus Protocol}
The I2C bus protocol coordinates the communication between multiple components in a master-slave hierarchy and is widely used in practice.
As it has no security features, this has led to exploits, for example, in smart cards of German public health insurance companies~\cite{I2C-article}.
The implementation used in this case study is taken from OpenCores~\cite{WEB:OpenCores}.
Its \aiger circuit consists of 254 latches plus 86 input and output variables.
Typically, this protocol consists of a master, one controller, and several slaves, where the master communicates to the slaves while the controller ensures properties like mutual exclusion.
Suppose information has to be sent over the bus, the master addresses the slaves with a designated address bit.
In this case study, we visualized the result of model checking the following information-flow policy: The information \emph{which} slave the master is addressing should not be identifiable from the bus' output.
This property is violated, but the counterexample is highly complex (e.g., it is not possible to generate a state graph).
Still, the visualizations provided by \hypervis help to understand the violation.

The analyst benefits especially from the highlighting and filtering functionalities.
The trace view of \hypervis is shown in \autoref{fig:i2c}.
\autoref{fig:i2c}a shows parts of the interface with highlighting activated.
While this already helps to see relevant elements, it is still nearly impossible to grasp the violation as large parts are outside of the browser's viewport.
By enabling the filtering and hiding variable types currently not of interest, see \autoref{fig:i2c}b, \hypervis can yield the proposition and the time step responsible for the violation and support the user in limiting the information to a reasonable amount.
Then, we can see that one output diverged at time step T6, resulting in a counterexample.

\paragraph{CS2: Symmetric Mutual Exclusion}
Arbiters form the basic building blocks in many protocols, such as the AMBA protocol or the above-mentioned I2C protocol.
Ensuring that no process or slave has an unfair advantage is highly desirable, also referred to as \emph{symmetry} in protocols.
In this case study, we visualize the results of checking if an arbiter implementation satisfies symmetry (cf.\ the arbiter system from \autoref{sec:challenges:workflows}).
Specifically, we check whether for two executions with symmetrically arriving requests the grants are also given symmetrically.

This case study features the interplay between the views implemented in \hypervis.
The \explanationView directly tells the analyst why the overall formula is violated.
When the highlighting button is pressed, \hypervis pinpoints the atomic propositions, time steps, and subformulas that caused the violation in the \traceView, the \timelineView, and the \formulaView (\autoref{fig:teaser}). 
In the \formulaView, for example, the subformula $grant\_0${\color{trace0}$@\pi$} $\leftrightarrow grant\_1${\color{trace1}$@\pi'$} is highlighted in the conclusion because only this subformula is needed to understand why the symmetry specification is violated: 
In the counterexample, $grant\_0${\color{trace0}$@\pi$} holds at time step $1$ while $grant\_1${\color{trace1}$@\pi'$} does not hold at that time step. 
Highlighting relevant subformulas decreases the number of subformulas that the analyst needs to consider when trying to understand the counterexample. 
This illustrates that \hypervis fulfills DG4, providing guidance for understanding the formula violation.

After the violation is identified, the bug in the system needs to be found.
Since DG3 is supported through the linked highlighting of elements and the highlight button, the \graphView is restricted to the relevant states for the counterexample executions. 
This feature again allows the analyst to focus their attention on the most relevant aspects.
By using \hypervis to explicitly step through the time steps, one observes that both executions represent the same system trace, thereby violating the symmetry in the grants. 
The solution to achieve symmetry is: Adding a new input to the system that allows giving grants symmetrically when both processes send requests simultaneously~\cite{MannaPnueli1996Safety}.

\subsection{User Feedback Sessions}
We conducted feedback sessions with domain users to better assess the merit of our tool for them.
In the following, we first describe the study design before reporting on the received feedback.

\subsubsection{Study Design}
\paragraph{Participants}
We recruited six participants (age M=\SI{27.5}{yrs}, SD=\SI{3.33}{yrs}; 1 female, 5 male) that have significant knowledge on model checking and hyperproperties.
On average, participants rated their theoretical expertise on model checking with 4.5 out 5 and on hyperproperties with 4.0.

\paragraph{Apparatus}
The sessions were conducted remotely through a video call with screen sharing.
We hosted the latest version of the tool online and provided participants with the link.
Two investigators moderated the videotaped sessions, and a third one was taking notes.
Participants were asked to follow a think-aloud protocol, i.e., verbally phrasing their thoughts and actions while interacting with the tool.

\paragraph{Procedure}
After a short welcome and general introduction, participants were asked to provide consent for the video recording.
Then, we outlined the procedure and think-aloud protocol before starting the demonstration of \hypervis via screen sharing, introducing all views and their functionalities.
Afterward, participants were asked to open the tool, start screen sharing from their end, and analyze three provided examples (detailed below).
For each example, we provided a short introduction on the specification and system and then asked them to reason about the counterexample.
Further, for the first example, participants had to propose a fix for a corrupt system, while in the second, they had to edit the formula to a working version.
For the last example, they had to identify the ``needle in the haystack''.
After working on each example, we asked them to reflect on the interface and which views they found helpful in the specific context.
Lastly, we concluded the session with an open discussion and provided them with a link to our questionnaire.
Sessions lasted one hour on average.

\paragraph{Provided Examples}
For the demonstration of \hypervis, we used an arbiter example similar to the one described in CS2.
Further, we prepared three examples for the hands-on part:
The first two consider a straightforward drone system that is supposed to increase the drone's height when it reads an $up$ input, and to go into an emergency state when a $bound$ input is read.
In the first version, the specification stated that equal $bound$ inputs must result in equal $emergency$ outputs in the next time step; however, the specification was violated due to an incorrect transition in the system.
Participants had to identify this issue and verbally provide a fix.
The counterexample is visible in \autoref{fig:views} and \autoref{fig:analysis}b.
In the second version, the fixed system was used, but now with a different but incorrect formula.
Participants had to pinpoint this issue and, this time, edit the formula in \hypervis.
The counterexample is shown in \autoref{fig:analysis}a.
Lastly, the third example was a larger counterexample involving 29 time steps and 50 atomic propositions, where a mutual exclusion specification was violated.
Due to the system's size, the \graphView was not available.
We asked participants to describe the violation in their own words and did not inquire any fixes.

\subsubsection{Results}
Overall, all participants (P1--P6) were able to work with \hypervis without larger issues
and considered the tool to be useful for experts and novices.
Our two main insights are: 
1) Our proposed interface allows to quickly identify the violations in counterexamples and provides valuable guidance for understanding the underlying issue.
2) Personal preferences and the different analysis workflows influence how the different views were used by the participants. 

All participants were able to correctly identify the violations and issues in the provided examples within the given time.
We could observe that the \emph{\traceView} was used as a central component within the analysis process, providing detailed information on the executions, while the \emph{linked highlighting} allowed for seeing the corresponding information in the other views.
For the \emph{\formulaView}, all participants (P1--P6) explicitly stated that they are intrigued by the hierarchy indicators below the formula.
Similarly, the \emph{\graphView} proved to be of great value, particularly when the traces were highlighted while stepping through the counterexample.
The comments and ratings also showed that the used \emph{colors for traces and statements} were appreciated for relating the different components.
Finally, the \emph{explanation highlighting} was considered ``extremely important'' (P3) for understanding the counterexample and identifying the relevant pieces for the violation (all Ps).

At the same time, not all views were used to the same extent.
For example, while most participants (P2--P6) found the \emph{textual explanations} ``extremely good'' (P6) and used it as starting point for understanding the violation, participant P1 preferred working with the other views.
Similarly, while participants P2+P5 only briefly used the \emph{stepping through mechanisms}, the others found it very helpful and used it extensively.
The \emph{\timelineView} was intensively utilized by participants P2 and P4--P6, 
particularly for comparing traces and recognizing specific patterns.
At the same time, P1 used the \traceView more extensively, while P3 used the \timelineView only for the larger example.

While working with \hypervis, participants also provided multiple suggestions for various improvements.
One commonly stated shortcoming was the missing \graphView for more extensive examples.
Further, P5+6 would prefer some indication of the explanation statements in the \graphView as well.
As participant P4 intensively used the stepping through functionalities, he proposed to improve the coloring of nodes and edges in the \graphView when both executions are overlapping.
Participants P4+6 suggested activating the explanation highlight on default.
Some extended filter options were also proposed, e.g., P5 suggested the filtering of single atomic propositions, while P2 proposed to allow for hiding time steps.
For the formula editing, multiple possible improvements were stated, e.g., better highlighting of corresponding brackets (P5+P6), a semantic check (P4+P5), or separating the \LaTeX{} input and rendered formula (P2).
Still, the formula editor was appreciated, with P4 stating that it is ``something that we needed for a long time''.

%

\section{Discussion}
The positive feedback that we received emphasizes that there is a clear need for visualization and analysis interfaces within the formal methods domain.
We found that the most important aspect when working on visualization solutions within this space is to have access to the specific knowledge that is involved in the rather abstract and formalized concepts.
From a visualization perspective, the incorporated encoding strategies or interactive mechanisms are mostly already known.
However, when applied and combined in the right way, they become extremely helpful.
Importantly, as it was also commented in our study, such a solution is not only an improvement for domain experts, but can also support novices in understanding the underlying principles.

Consequently, our work is in line with other efforts of providing explications and intuition for abstract or black-box-like processes~\cite{Strobelt2019Seq2seq, Flemisch2020Towards, Wiehr2020SafeHandoverMixed}.
However, in this area, work around explainable AI~\cite{Gunnning2016Explainable} has received most attention in recent years, while formal methods themselves are only rarely considered.
This is particularly interesting for two reasons:
(1) formal methods, and especially model checking, are largely built around mathematical and logical representations that are consumed in command-line tools, while visual representations remain underestimated.
Therefore, there is a big potential for making the concepts more accessible by using visualizations.
Further, the mathematical nature of it requires a rigorous treatment of the visualized elements, which poses special challenges to the visualization design.
(2) Formal methods also play an important role for AI in general and when trying to provide explanations to computations of an AI agent.
For example, Marabou~\cite{DBLP:conf/cav/KatzHIJLLSTWZDK19} is a recently introduced framework for verifying and providing counterexamples of properties of deep neural networks (e.g., robustness, which is in general a hyperproperty).
However, as it is command-line based and does not provide an explanation on the counterexample, users have to cope with the same problems described in this work.
The here presented visualization approaches might be directly applicable to many tools in the area of formal methods.

In the light of these considerations, \hypervis can be seen as a first foundation for explaining hyperproperties and counterexamples.
As the immediate next steps, the suggested improvements from the user feedback sessions can be incorporated.
For the editing facilities, this includes the general possibility for modifying the system plus a visual editing mode.
This could also allow for providing a stand-alone editing mode with improved live previews of formula and system.
For the analysis of counterexamples themselves, an interesting addition would be support for adding annotations or storing derived insights~\cite{Mathisen2019InsideInsightsIntegratingData}.
In this context, it can also be considered to automatically track the analysis history or provenance~\cite{Xu2020SurveyAnalysisUser} and allow analysts to review it.

Currently, \hypervis is focused on visualizing a specific counterexample to a hyperproperty.
However, on the one hand, the need for visually representing hyperproperties can also occur independently from violated specifications, i.e., for correctly implemented systems and specified properties.
While it is always possible to generate a so-called witness for a proved hyperproperty by negating the specification, the found witness is one of many possible ones and might not adequately represent the underlying hyperproperty.
On the other hand, the challenging but promising interactive synthesis problem potentially benefits from the presented visualizations.
Synthesis constructs per definition a correct implementation directly from the provided specification, making the model checking process superfluous.
Here, it would be beneficial to visualize the iterative synthesis process (i.e., how the system was derived) as well as the proposed implementation itself.
The visualization and interaction designs presented here can guide the development of such novel hyperproperty visualization tools.

%

\section{Conclusion}
In this paper, we presented concepts for visually analyzing counterexamples to hyperproperties as well as for editing the provided formula and system.
As demonstrated through case studies and attested by user feedback, our \hypervis tool notably improves the analysis and understanding of the counterexamples.
At the core of this is the targeted usage of encoding strategies and interactive mechanisms that pointedly represent the different aspects and help to guide the analyst to the relevant information in the example.
In particular, the right combination of allegedly simple measures, such as color encoding, linked highlighting, and relevance indication, can allow experts to quickly recognize the violation cause and also novices to understand the complex relations in the first place.
Notably, the key to such solutions is the understanding of the domain, which in this case enabled us to embed the causal analysis of the counterexample and to automatically derive textual explanations and corresponding highlights.
The provided editing facilities support fixing the identified issues, turning \hypervis into a valuable tool for analyzing hyperproperties.
With this, we contribute a foundation for explaining and visualizing hyperproperties in general and hope to inspire further visualization solutions for more formal methods concepts.

\acknowledgments{
We thank Weizhou Luo for his valuable support during the overall project duration.
This work was funded by DFG grant 389792660 as part of \href{https://perspicuous-computing.science}{TRR~248 -- CPEC}, by the DFG as part of the Germany’s Excellence Strategy EXC 2050/1 - Project ID 390696704  - Cluster of Excellence ``\emph{Centre for Tactile Internet}'' (CeTI) of TU Dresden, by the European Research Council (ERC)\ Grant OSARES (No. 683300), and by the German Israeli Foundation (GIF) Grant No. I-1513-407./2019.
}

\bibliographystyle{abbrv-doi-hyperref-narrow}

\bibliography{refs}

\begin{thebibliography}{10}
\renewcommand*{\sfdefault}{PTSansNarrow-TLF}

\bibitem{FSM-to-VHDL}
\href{https://doi.org/10.1109/CCECE.2004.1347584}{A.~T. {Abdel-Hamid},
  M.~{Zaki}, and S.~{Tahar}}.
\newblock \href{https://doi.org/10.1109/CCECE.2004.1347584}{A tool converting
  finite state machine to vhdl}.
\newblock \href{https://doi.org/10.1109/CCECE.2004.1347584}{In {\em Canadian
  Conference on Electrical and Computer Engineering 2004}},
  \href{https://doi.org/10.1109/CCECE.2004.1347584}{pp. 1907--1910},
  \href{https://doi.org/10.1109/CCECE.2004.1347584}{2004}.
  \href{https://doi.org/10.1109/CCECE.2004.1347584}
{doi: \textsf{%
10\hspace{.1pt}\discretionary{.}{%
}{.}\hspace{.4pt}1109\discretionary{/}{%
}{/}CCECE\hspace{.1pt}\discretionary{.}{%
}{.}\hspace{.4pt}2004\hspace{.1pt}\discretionary{.}{%
}{.}\hspace{.4pt}1347584}}


\bibitem{Abrahamson2014SheddingLightDistributed}
\href{https://doi.org/10.1145/2591062.2591134}{J.~Abrahamson, I.~Beschastnikh,
  Y.~Brun, and M.~D. Ernst}.
\newblock \href{https://doi.org/10.1145/2591062.2591134}{Shedding light on
  distributed system executions}.
\newblock \href{https://doi.org/10.1145/2591062.2591134}{In {\em Companion
  Proc. International Conference on Software Engineering}},
  \href{https://doi.org/10.1145/2591062.2591134}{pp. 598--599}.
  \href{https://doi.org/10.1145/2591062.2591134}{{ACM}},
  \href{https://doi.org/10.1145/2591062.2591134}{New York, NY, USA},
  \href{https://doi.org/10.1145/2591062.2591134}{2014}.
  \href{https://doi.org/10.1145/2591062.2591134}
{doi: \textsf{%
10\hspace{.1pt}\discretionary{.}{%
}{.}\hspace{.4pt}1145\discretionary{/}{%
}{/}2591062\hspace{.1pt}\discretionary{.}{%
}{.}\hspace{.4pt}2591134}}


\bibitem{Aljazzar2008DebuggingDependabilityModels}
\href{https://doi.org/10.1109/qest.2008.40}{H.~Aljazzar and S.~Leue}.
\newblock \href{https://doi.org/10.1109/qest.2008.40}{Debugging of
  dependability models using interactive visualization of counterexamples}.
\newblock \href{https://doi.org/10.1109/qest.2008.40}{In {\em International
  Conference on Quantitative Evaluation of Systems}},
  \href{https://doi.org/10.1109/qest.2008.40}{pp. 189--198}.
  \href{https://doi.org/10.1109/qest.2008.40}{{IEEE}},
  \href{https://doi.org/10.1109/qest.2008.40}{Piscataway, NJ, USA},
  \href{https://doi.org/10.1109/qest.2008.40}{2008}.
  \href{https://doi.org/10.1109/qest.2008.40}
{doi: \textsf{%
10\hspace{.1pt}\discretionary{.}{%
}{.}\hspace{.4pt}1109\discretionary{/}{%
}{/}qest\hspace{.1pt}\discretionary{.}{%
}{.}\hspace{.4pt}2008\hspace{.1pt}\discretionary{.}{%
}{.}\hspace{.4pt}40}}


\bibitem{Simulink}
\href{https://doi.org/doi:10.1515/9783110636420}{A.~Angermann, M.~Beuschel,
  M.~Rau, and U.~Wohlfarth}.
\newblock \href{https://doi.org/doi:10.1515/9783110636420}{{\em MATLAB –
  Simulink – Stateflow}}.
\newblock \href{https://doi.org/doi:10.1515/9783110636420}{De Gruyter
  Oldenbourg}, \href{https://doi.org/doi:10.1515/9783110636420}{2020}.
  \href{https://doi.org/doi:10.1515/9783110636420}
{doi: \textsf{%
doi\discretionary{:}{%
}{:}10\hspace{.1pt}\discretionary{.}{%
}{.}\hspace{.4pt}1515\discretionary{/}{%
}{/}9783110636420}}


\bibitem{Barthe2011SecureInformationFlow}
\href{https://doi.org/10.1017/s0960129511000193}{G.~Barthe, P.~R.
  D{\textquotesingle}Argenio, and T.~Rezk}.
\newblock \href{https://doi.org/10.1017/s0960129511000193}{Secure information
  flow by self-composition}.
\newblock \href{https://doi.org/10.1017/s0960129511000193}{{\em Mathematical
  Structures in Computer Science}},
  \href{https://doi.org/10.1017/s0960129511000193}{21(6):1207--1252},
  \href{https://doi.org/10.1017/s0960129511000193}{Oct. 2011}.
  \href{https://doi.org/10.1017/s0960129511000193}
{doi: \textsf{%
10\hspace{.1pt}\discretionary{.}{%
}{.}\hspace{.4pt}1017\discretionary{/}{%
}{/}s0960129511000193}}


\bibitem{Beck2017WordSizedGraphics}
\href{https://doi.org/10.1109/tvcg.2017.2674958}{F.~Beck and D.~Weiskopf}.
\newblock \href{https://doi.org/10.1109/tvcg.2017.2674958}{Word-sized graphics
  for scientific texts}.
\newblock \href{https://doi.org/10.1109/tvcg.2017.2674958}{{\em {{IEEE} Trans.
  Visualization and Computer Graphics}}},
  \href{https://doi.org/10.1109/tvcg.2017.2674958}{23(6):1576--1587},
  \href{https://doi.org/10.1109/tvcg.2017.2674958}{June 2017}.
  \href{https://doi.org/10.1109/tvcg.2017.2674958}
{doi: \textsf{%
10\hspace{.1pt}\discretionary{.}{%
}{.}\hspace{.4pt}1109\discretionary{/}{%
}{/}tvcg\hspace{.1pt}\discretionary{.}{%
}{.}\hspace{.4pt}2017\hspace{.1pt}\discretionary{.}{%
}{.}\hspace{.4pt}2674958}}


\bibitem{Beer2009ExplainingCounterexamplesUsing}
\href{https://doi.org/10.1007/978-3-642-02658-4_11}{I.~Beer, S.~Ben-David,
  H.~Chockler, A.~Orni, and R.~Trefler}.
\newblock \href{https://doi.org/10.1007/978-3-642-02658-4_11}{Explaining
  counterexamples using causality}.
\newblock \href{https://doi.org/10.1007/978-3-642-02658-4_11}{In {\em Computer
  Aided Verification}}, \href{https://doi.org/10.1007/978-3-642-02658-4_11}{pp.
  94--108}. \href{https://doi.org/10.1007/978-3-642-02658-4_11}{Springer Berlin
  Heidelberg}, \href{https://doi.org/10.1007/978-3-642-02658-4_11}{2009}.
  \href{https://doi.org/10.1007/978-3-642-02658-4_11}
{doi: \textsf{%
10\hspace{.1pt}\discretionary{.}{%
}{.}\hspace{.4pt}1007\discretionary{/}{%
}{/}978\discretionary{%
}{-}{-}3\discretionary{%
}{-}{-}642\discretionary{%
}{-}{-}02658\discretionary{%
}{-}{-}4\_11}}


\bibitem{Beschastnikh2020VisualizingDistributedSystem}
\href{https://doi.org/10.1145/3375633}{I.~Beschastnikh, P.~Liu, A.~Xing,
  P.~Wang, Y.~Brun, and M.~D. Ernst}.
\newblock \href{https://doi.org/10.1145/3375633}{Visualizing distributed system
  executions}.
\newblock \href{https://doi.org/10.1145/3375633}{{\em {ACM} Transactions on
  Software Engineering and Methodology}},
  \href{https://doi.org/10.1145/3375633}{29(2):9:1--9:38},
  \href{https://doi.org/10.1145/3375633}{Apr. 2020}.
  \href{https://doi.org/10.1145/3375633}
{doi: \textsf{%
10\hspace{.1pt}\discretionary{.}{%
}{.}\hspace{.4pt}1145\discretionary{/}{%
}{/}3375633}}


\bibitem{Biere2007AIGERInverterGraph}
A.~Biere.
\newblock The {AIGER And-Inverter Graph (AIG)} format version 20071012.
\newblock Technical Report Report 07/1, Institute for Formal Models and
  Verification, Johannes Kepler University, Linz, Austria, 2007.

\bibitem{Biere2011AIGER1.9}
A.~Biere, K.~Heljanko, and S.~Wieringa.
\newblock Aiger 1.9 and beyond.
\newblock Technical report, Institute for Formal Models and Verification,
  Johannes Kepler University, Linz, Austria, 2011.

\bibitem{Bochot2010PathsPropertyViolation}
\href{https://doi.org/10.1109/hase.2010.15}{T.~Bochot, P.~Virelizier,
  H.~Waeselynck, and V.~Wiels}.
\newblock \href{https://doi.org/10.1109/hase.2010.15}{Paths to property
  violation: A structural approach for analyzing counter-examples}.
\newblock \href{https://doi.org/10.1109/hase.2010.15}{In {\em {IEEE}
  International Symposium on High Assurance Systems Engineering}},
  \href{https://doi.org/10.1109/hase.2010.15}{pp. 74--83}.
  \href{https://doi.org/10.1109/hase.2010.15}{{IEEE}},
  \href{https://doi.org/10.1109/hase.2010.15}{Piscataway, NJ, USA},
  \href{https://doi.org/10.1109/hase.2010.15}{2010}.
  \href{https://doi.org/10.1109/hase.2010.15}
{doi: \textsf{%
10\hspace{.1pt}\discretionary{.}{%
}{.}\hspace{.4pt}1109\discretionary{/}{%
}{/}hase\hspace{.1pt}\discretionary{.}{%
}{.}\hspace{.4pt}2010\hspace{.1pt}\discretionary{.}{%
}{.}\hspace{.4pt}15}}


\bibitem{Bolton2010UsingTaskAnalytic}
\href{https://doi.org/10.1109/icsmc.2010.5641711}{M.~L. Bolton and E.~J. Bass}.
\newblock \href{https://doi.org/10.1109/icsmc.2010.5641711}{Using task analytic
  models to visualize model checker counterexamples}.
\newblock \href{https://doi.org/10.1109/icsmc.2010.5641711}{In {\em Intl.
  Conference on Systems, Man and Cybernetics}},
  \href{https://doi.org/10.1109/icsmc.2010.5641711}{pp. 2069--2074}.
  \href{https://doi.org/10.1109/icsmc.2010.5641711}{{IEEE}},
  \href{https://doi.org/10.1109/icsmc.2010.5641711}{2010}.
  \href{https://doi.org/10.1109/icsmc.2010.5641711}
{doi: \textsf{%
10\hspace{.1pt}\discretionary{.}{%
}{.}\hspace{.4pt}1109\discretionary{/}{%
}{/}icsmc\hspace{.1pt}\discretionary{.}{%
}{.}\hspace{.4pt}2010\hspace{.1pt}\discretionary{.}{%
}{.}\hspace{.4pt}5641711}}


\bibitem{Bostock2011D3DataDriven}
\href{https://doi.org/10.1109/TVCG.2011.185}{M.~Bostock, V.~Ogievetsky, and
  J.~Heer}.
\newblock \href{https://doi.org/10.1109/TVCG.2011.185}{D$^3$: Data-driven
  documents}.
\newblock \href{https://doi.org/10.1109/TVCG.2011.185}{{\em {{IEEE} Trans.
  Visualization and Computer Graphics}}},
  \href{https://doi.org/10.1109/TVCG.2011.185}{17(12):2301--2309},
  \href{https://doi.org/10.1109/TVCG.2011.185}{Dec. 2011}.
  \href{https://doi.org/10.1109/TVCG.2011.185}
{doi: \textsf{%
10\hspace{.1pt}\discretionary{.}{%
}{.}\hspace{.4pt}1109\discretionary{/}{%
}{/}TVCG\hspace{.1pt}\discretionary{.}{%
}{.}\hspace{.4pt}2011\hspace{.1pt}\discretionary{.}{%
}{.}\hspace{.4pt}185}}


\bibitem{Brayton2010ABCAcademicIndustrial}
\href{https://doi.org/10.1007/978-3-642-14295-6_5}{R.~Brayton and
  A.~Mishchenko}.
\newblock \href{https://doi.org/10.1007/978-3-642-14295-6_5}{{ABC}: An academic
  industrial-strength verification tool}.
\newblock \href{https://doi.org/10.1007/978-3-642-14295-6_5}{In {\em Computer
  Aided Verification}}, \href{https://doi.org/10.1007/978-3-642-14295-6_5}{pp.
  24--40}. \href{https://doi.org/10.1007/978-3-642-14295-6_5}{Springer Berlin
  Heidelberg}, \href{https://doi.org/10.1007/978-3-642-14295-6_5}{2010}.
  \href{https://doi.org/10.1007/978-3-642-14295-6_5}
{doi: \textsf{%
10\hspace{.1pt}\discretionary{.}{%
}{.}\hspace{.4pt}1007\discretionary{/}{%
}{/}978\discretionary{%
}{-}{-}3\discretionary{%
}{-}{-}642\discretionary{%
}{-}{-}14295\discretionary{%
}{-}{-}6\_5}}


\bibitem{Buja1991InteractiveDataVisualization}
\href{https://doi.org/10.1109/visual.1991.175794}{A.~Buja, J.~McDonald,
  J.~Michalak, and W.~Stuetzle}.
\newblock \href{https://doi.org/10.1109/visual.1991.175794}{Interactive data
  visualization using focusing and linking}.
\newblock \href{https://doi.org/10.1109/visual.1991.175794}{In {\em Proc.
  Visualization {\textquotesingle}91}},
  \href{https://doi.org/10.1109/visual.1991.175794}{pp. 156--163}.
  \href{https://doi.org/10.1109/visual.1991.175794}{{IEEE}},
  \href{https://doi.org/10.1109/visual.1991.175794}{Piscataway, NJ, USA},
  \href{https://doi.org/10.1109/visual.1991.175794}{1991}.
  \href{https://doi.org/10.1109/visual.1991.175794}
{doi: \textsf{%
10\hspace{.1pt}\discretionary{.}{%
}{.}\hspace{.4pt}1109\discretionary{/}{%
}{/}visual\hspace{.1pt}\discretionary{.}{%
}{.}\hspace{.4pt}1991\hspace{.1pt}\discretionary{.}{%
}{.}\hspace{.4pt}175794}}


\bibitem{Campetelli2011UserfriendlyModel}
A.~Campetelli, F.~H{\"o}lzl, and P.~Neubeck.
\newblock User-friendly model checking integration in model-based development.
\newblock In {\em International Conference on Computer Applications in Industry
  and Engineering}, 2011.

\bibitem{MathMLV3-2014}
D.~Carlisle, P.~D.~F. Ion, and R.~R. Miner.
\newblock {\em Mathematical Markup Language (MathML) Version 3.0 2nd Edition}.
\newblock {W3C}, 2014.

\bibitem{cervone2015towards}
D.~Cervone, P.~Krautzberger, and V.~Sorge.
\newblock Towards meaningful visual abstraction of mathematical notation.
\newblock {\em Proc. CICM}, 2015.

\bibitem{Chaki2004ModularVerificationSoftware}
\href{https://doi.org/10.1109/tse.2004.22}{S.~Chaki, E.~Clarke, A.~Groce,
  S.~Jha, and H.~Veith}.
\newblock \href{https://doi.org/10.1109/tse.2004.22}{Modular verification of
  software components in c}.
\newblock \href{https://doi.org/10.1109/tse.2004.22}{{\em {IEEE} Transactions
  on Software Engineering}},
  \href{https://doi.org/10.1109/tse.2004.22}{30(6):388--402},
  \href{https://doi.org/10.1109/tse.2004.22}{June 2004}.
  \href{https://doi.org/10.1109/tse.2004.22}
{doi: \textsf{%
10\hspace{.1pt}\discretionary{.}{%
}{.}\hspace{.4pt}1109\discretionary{/}{%
}{/}tse\hspace{.1pt}\discretionary{.}{%
}{.}\hspace{.4pt}2004\hspace{.1pt}\discretionary{.}{%
}{.}\hspace{.4pt}22}}


\bibitem{Chaki2004ExplainingAbstractCounterexamples}
\href{https://doi.org/10.1145/1029894.1029908}{S.~Chaki, A.~Groce, and
  O.~Strichman}.
\newblock \href{https://doi.org/10.1145/1029894.1029908}{Explaining abstract
  counterexamples}.
\newblock \href{https://doi.org/10.1145/1029894.1029908}{In {\em Proc. Intl.
  Symposium on Foundations of Software Engineering}},
  \href{https://doi.org/10.1145/1029894.1029908}{pp. 73--–82}.
  \href{https://doi.org/10.1145/1029894.1029908}{{ACM}},
  \href{https://doi.org/10.1145/1029894.1029908}{New York, NY, USA},
  \href{https://doi.org/10.1145/1029894.1029908}{2004}.
  \href{https://doi.org/10.1145/1029894.1029908}
{doi: \textsf{%
10\hspace{.1pt}\discretionary{.}{%
}{.}\hspace{.4pt}1145\discretionary{/}{%
}{/}1029894\hspace{.1pt}\discretionary{.}{%
}{.}\hspace{.4pt}1029908}}


\bibitem{Chen2021CompositionConfigurationPatterns}
\href{https://doi.org/10.1109/tvcg.2020.3030338}{X.~Chen, W.~Zeng, Y.~Lin,
  H.~M. Al-maneea, J.~Roberts, and R.~Chang}.
\newblock \href{https://doi.org/10.1109/tvcg.2020.3030338}{Composition and
  configuration patterns in multiple-view visualizations}.
\newblock \href{https://doi.org/10.1109/tvcg.2020.3030338}{{\em {{IEEE} Trans.
  Visualization and Computer Graphics}}},
  \href{https://doi.org/10.1109/tvcg.2020.3030338}{27(2):1514--1524},
  \href{https://doi.org/10.1109/tvcg.2020.3030338}{Feb. 2021}.
  \href{https://doi.org/10.1109/tvcg.2020.3030338}
{doi: \textsf{%
10\hspace{.1pt}\discretionary{.}{%
}{.}\hspace{.4pt}1109\discretionary{/}{%
}{/}tvcg\hspace{.1pt}\discretionary{.}{%
}{.}\hspace{.4pt}2020\hspace{.1pt}\discretionary{.}{%
}{.}\hspace{.4pt}3030338}}


\bibitem{Clarke2004ToolCheckingANSI}
\href{https://doi.org/10.1007/978-3-540-24730-2_15}{E.~Clarke, D.~Kroening, and
  F.~Lerda}.
\newblock \href{https://doi.org/10.1007/978-3-540-24730-2_15}{A tool for
  checking {ANSI}-c programs}.
\newblock \href{https://doi.org/10.1007/978-3-540-24730-2_15}{In {\em Tools and
  Algorithms for the Construction and Analysis of Systems}},
  \href{https://doi.org/10.1007/978-3-540-24730-2_15}{pp. 168--176}.
  \href{https://doi.org/10.1007/978-3-540-24730-2_15}{Springer Berlin
  Heidelberg}, \href{https://doi.org/10.1007/978-3-540-24730-2_15}{2004}.
  \href{https://doi.org/10.1007/978-3-540-24730-2_15}
{doi: \textsf{%
10\hspace{.1pt}\discretionary{.}{%
}{.}\hspace{.4pt}1007\discretionary{/}{%
}{/}978\discretionary{%
}{-}{-}3\discretionary{%
}{-}{-}540\discretionary{%
}{-}{-}24730\discretionary{%
}{-}{-}2\_15}}


\bibitem{10.5555/332656}
E.~M. Clarke, O.~Grumberg, and D.~A. Peled.
\newblock {\em Model Checking}.
\newblock MIT Press, Cambridge, MA, USA, 2000.

\bibitem{Clarkson2014TemporalLogicsHyperproperties}
\href{https://doi.org/10.1007/978-3-642-54792-8_15}{M.~R. Clarkson,
  B.~Finkbeiner, M.~Koleini, K.~K. Micinski, M.~N. Rabe, and C.~S{\'{a}}nchez}.
\newblock \href{https://doi.org/10.1007/978-3-642-54792-8_15}{Temporal logics
  for hyperproperties}.
\newblock \href{https://doi.org/10.1007/978-3-642-54792-8_15}{In {\em Lecture
  Notes in Computer Science}},
  \href{https://doi.org/10.1007/978-3-642-54792-8_15}{pp. 265--284}.
  \href{https://doi.org/10.1007/978-3-642-54792-8_15}{Springer Berlin
  Heidelberg}, \href{https://doi.org/10.1007/978-3-642-54792-8_15}{2014}.
  \href{https://doi.org/10.1007/978-3-642-54792-8_15}
{doi: \textsf{%
10\hspace{.1pt}\discretionary{.}{%
}{.}\hspace{.4pt}1007\discretionary{/}{%
}{/}978\discretionary{%
}{-}{-}3\discretionary{%
}{-}{-}642\discretionary{%
}{-}{-}54792\discretionary{%
}{-}{-}8\_15}}


\bibitem{Coenen2019HierarchyHyperlogics}
\href{https://doi.org/10.1109/lics.2019.8785713}{N.~Coenen, B.~Finkbeiner,
  C.~Hahn, and J.~Hofmann}.
\newblock \href{https://doi.org/10.1109/lics.2019.8785713}{The hierarchy of
  hyperlogics}.
\newblock \href{https://doi.org/10.1109/lics.2019.8785713}{In {\em Annual
  {ACM}/{IEEE} Symposium on Logic in Computer Science}}.
  \href{https://doi.org/10.1109/lics.2019.8785713}{{IEEE}},
  \href{https://doi.org/10.1109/lics.2019.8785713}{June 2019}.
  \href{https://doi.org/10.1109/lics.2019.8785713}
{doi: \textsf{%
10\hspace{.1pt}\discretionary{.}{%
}{.}\hspace{.4pt}1109\discretionary{/}{%
}{/}lics\hspace{.1pt}\discretionary{.}{%
}{.}\hspace{.4pt}2019\hspace{.1pt}\discretionary{.}{%
}{.}\hspace{.4pt}8785713}}


\bibitem{DBLP:conf/cav/CoenenFST19}
\href{https://doi.org/10.1007/978-3-030-25540-4_7}{N.~Coenen, B.~Finkbeiner,
  C.~S{\'{a}}nchez, and L.~Tentrup}.
\newblock \href{https://doi.org/10.1007/978-3-030-25540-4_7}{Verifying
  hyperliveness}.
\newblock \href{https://doi.org/10.1007/978-3-030-25540-4_7}{In {\em Intl.
  Conference on Computer Aided Verification}},
  \href{https://doi.org/10.1007/978-3-030-25540-4_7}{pp. 121--139}.
  \href{https://doi.org/10.1007/978-3-030-25540-4_7}{Springer},
  \href{https://doi.org/10.1007/978-3-030-25540-4_7}{2019}.
  \href{https://doi.org/10.1007/978-3-030-25540-4_7}
{doi: \textsf{%
10\hspace{.1pt}\discretionary{.}{%
}{.}\hspace{.4pt}1007\discretionary{/}{%
}{/}978\discretionary{%
}{-}{-}3\discretionary{%
}{-}{-}030\discretionary{%
}{-}{-}25540\discretionary{%
}{-}{-}4\_7}}


\bibitem{DBLP:conf/cav/CookKKTTT18}
\href{https://doi.org/10.1007/978-3-319-96142-2_28}{B.~Cook, K.~Khazem,
  D.~Kroening, S.~Tasiran, M.~Tautschnig, and M.~R. Tuttle}.
\newblock \href{https://doi.org/10.1007/978-3-319-96142-2_28}{Model checking
  boot code from {AWS} data centers}.
\newblock \href{https://doi.org/10.1007/978-3-319-96142-2_28}{In {\em
  International Conference on Computer Aided Verification}},
  \href{https://doi.org/10.1007/978-3-319-96142-2_28}{pp. 467--486}.
  \href{https://doi.org/10.1007/978-3-319-96142-2_28}{Springer},
  \href{https://doi.org/10.1007/978-3-319-96142-2_28}{2018}.
  \href{https://doi.org/10.1007/978-3-319-96142-2_28}
{doi: \textsf{%
10\hspace{.1pt}\discretionary{.}{%
}{.}\hspace{.4pt}1007\discretionary{/}{%
}{/}978\discretionary{%
}{-}{-}3\discretionary{%
}{-}{-}319\discretionary{%
}{-}{-}96142\discretionary{%
}{-}{-}2\_28}}


\bibitem{Cui2019DataSite}
\href{https://doi.org/10.1177/1473871618806555}{Z.~Cui, S.~K. Badam, M.~A.
  Yalçin, and N.~Elmqvist}.
\newblock \href{https://doi.org/10.1177/1473871618806555}{{DataSite}: Proactive
  visual data exploration with computation of insight-based recommendations}.
\newblock \href{https://doi.org/10.1177/1473871618806555}{{\em Information
  Visualization}},
  \href{https://doi.org/10.1177/1473871618806555}{18(2):251--267},
  \href{https://doi.org/10.1177/1473871618806555}{2019}.
  \href{https://doi.org/10.1177/1473871618806555}
{doi: \textsf{%
10\hspace{.1pt}\discretionary{.}{%
}{.}\hspace{.4pt}1177\discretionary{/}{%
}{/}1473871618806555}}


\bibitem{DeepNeuralNetHandwritingRecognition}
\href{https://doi.org/10.1109/ACPR.2015.7486478}{H.~{Dai Nguyen}, A.~D. {Le},
  and M.~{Nakagawa}}.
\newblock \href{https://doi.org/10.1109/ACPR.2015.7486478}{Deep neural networks
  for recognizing online handwritten mathematical symbols}.
\newblock \href{https://doi.org/10.1109/ACPR.2015.7486478}{In {\em IAPR Asian
  Conference on Pattern Recognition}},
  \href{https://doi.org/10.1109/ACPR.2015.7486478}{pp. 121--125},
  \href{https://doi.org/10.1109/ACPR.2015.7486478}{2015}.
  \href{https://doi.org/10.1109/ACPR.2015.7486478}
{doi: \textsf{%
10\hspace{.1pt}\discretionary{.}{%
}{.}\hspace{.4pt}1109\discretionary{/}{%
}{/}ACPR\hspace{.1pt}\discretionary{.}{%
}{.}\hspace{.4pt}2015\hspace{.1pt}\discretionary{.}{%
}{.}\hspace{.4pt}7486478}}


\bibitem{DBLP:conf/cav/FinkbeinerHT18}
\href{https://doi.org/10.1007/978-3-319-96145-3_8}{B.~Finkbeiner, C.~Hahn, and
  H.~Torfah}.
\newblock \href{https://doi.org/10.1007/978-3-319-96145-3_8}{Model checking
  quantitative hyperproperties}.
\newblock \href{https://doi.org/10.1007/978-3-319-96145-3_8}{In {\em Intl.
  Conference on Computer Aided Verification}},
  \href{https://doi.org/10.1007/978-3-319-96145-3_8}{pp. 144--163}.
  \href{https://doi.org/10.1007/978-3-319-96145-3_8}{Springer},
  \href{https://doi.org/10.1007/978-3-319-96145-3_8}{2018}.
  \href{https://doi.org/10.1007/978-3-319-96145-3_8}
{doi: \textsf{%
10\hspace{.1pt}\discretionary{.}{%
}{.}\hspace{.4pt}1007\discretionary{/}{%
}{/}978\discretionary{%
}{-}{-}3\discretionary{%
}{-}{-}319\discretionary{%
}{-}{-}96145\discretionary{%
}{-}{-}3\_8}}


\bibitem{Finkbeiner2015AlgorithmsModelChecking}
\href{https://doi.org/10.1007/978-3-319-21690-4_3}{B.~Finkbeiner, M.~N. Rabe,
  and C.~S{\'{a}}nchez}.
\newblock \href{https://doi.org/10.1007/978-3-319-21690-4_3}{Algorithms for
  model checking {HyperLTL} and {HyperCTL$^*$}}.
\newblock \href{https://doi.org/10.1007/978-3-319-21690-4_3}{In {\em Computer
  Aided Verification}},
  \href{https://doi.org/10.1007/978-3-319-21690-4_3}{Lecture Notes in Computer
  Science}, \href{https://doi.org/10.1007/978-3-319-21690-4_3}{pp. 30--48}.
  \href{https://doi.org/10.1007/978-3-319-21690-4_3}{Springer International
  Publishing}, \href{https://doi.org/10.1007/978-3-319-21690-4_3}{2015}.
  \href{https://doi.org/10.1007/978-3-319-21690-4_3}
{doi: \textsf{%
10\hspace{.1pt}\discretionary{.}{%
}{.}\hspace{.4pt}1007\discretionary{/}{%
}{/}978\discretionary{%
}{-}{-}3\discretionary{%
}{-}{-}319\discretionary{%
}{-}{-}21690\discretionary{%
}{-}{-}4\_3}}


\bibitem{Flemisch2020Towards}
T.~Flemisch, R.~Langner, C.~Alrabbaa, and R.~Dachselt.
\newblock Towards designing a tool for understanding proofs in ontologies
  through combined node-link diagrams.
\newblock In {\em International Workshop on Visualization and Interaction for
  Ontologies and Linked Data}, Nov. 2020.

\bibitem{Frisch2012Node-link-editing}
\href{http://www.dr.hut-verlag.de/978-3-8439-0563-3.html}{M.~Frisch}.
\newblock \href{http://www.dr.hut-verlag.de/978-3-8439-0563-3.html}{{\em
  Visualization and Interaction Techniques for Node-Link Diagram Editing and
  Exploration}}.
\newblock \href{http://www.dr.hut-verlag.de/978-3-8439-0563-3.html}{PhD
  thesis},
  \href{http://www.dr.hut-verlag.de/978-3-8439-0563-3.html}{Otto-von-Guericke-Universit\"{a}t
  Magdeburg},
  \href{http://www.dr.hut-verlag.de/978-3-8439-0563-3.html}{M\"{u}nchen},
  \href{http://www.dr.hut-verlag.de/978-3-8439-0563-3.html}{6 2012}.

\bibitem{Gleiss2019InteractiveVisualizationSaturation}
\href{https://doi.org/10.1007/978-3-030-34968-4_28}{B.~Gleiss, L.~Kov{\'{a}}cs,
  and L.~Schnedlitz}.
\newblock \href{https://doi.org/10.1007/978-3-030-34968-4_28}{Interactive
  visualization of saturation attempts in vampire}.
\newblock \href{https://doi.org/10.1007/978-3-030-34968-4_28}{In {\em Lecture
  Notes in Computer Science}},
  \href{https://doi.org/10.1007/978-3-030-34968-4_28}{pp. 504--513}.
  \href{https://doi.org/10.1007/978-3-030-34968-4_28}{Springer International
  Publishing}, \href{https://doi.org/10.1007/978-3-030-34968-4_28}{2019}.
  \href{https://doi.org/10.1007/978-3-030-34968-4_28}
{doi: \textsf{%
10\hspace{.1pt}\discretionary{.}{%
}{.}\hspace{.4pt}1007\discretionary{/}{%
}{/}978\discretionary{%
}{-}{-}3\discretionary{%
}{-}{-}030\discretionary{%
}{-}{-}34968\discretionary{%
}{-}{-}4\_28}}


\bibitem{Goffin2017ExploratoryStudyWord}
\href{https://doi.org/10.1109/tvcg.2016.2618797}{P.~Goffin, J.~Boy, W.~Willett,
  and P.~Isenberg}.
\newblock \href{https://doi.org/10.1109/tvcg.2016.2618797}{An exploratory study
  of word-scale graphics in data-rich text documents}.
\newblock \href{https://doi.org/10.1109/tvcg.2016.2618797}{{\em {{IEEE} Trans.
  Visualization and Computer Graphics}}},
  \href{https://doi.org/10.1109/tvcg.2016.2618797}{23(10):2275--2287},
  \href{https://doi.org/10.1109/tvcg.2016.2618797}{Oct. 2017}.
  \href{https://doi.org/10.1109/tvcg.2016.2618797}
{doi: \textsf{%
10\hspace{.1pt}\discretionary{.}{%
}{.}\hspace{.4pt}1109\discretionary{/}{%
}{/}tvcg\hspace{.1pt}\discretionary{.}{%
}{.}\hspace{.4pt}2016\hspace{.1pt}\discretionary{.}{%
}{.}\hspace{.4pt}2618797}}


\bibitem{Goldsby2006VisualizationFrameworkModeling}
\href{https://doi.org/10.1007/11880240_49}{H.~Goldsby, B.~H.~C. Cheng,
  S.~Konrad, and S.~Kamdoum}.
\newblock \href{https://doi.org/10.1007/11880240_49}{A visualization framework
  for the modeling and formal analysis of high assurance systems}.
\newblock \href{https://doi.org/10.1007/11880240_49}{In {\em Model Driven
  Engineering Languages and Systems}},
  \href{https://doi.org/10.1007/11880240_49}{pp. 707--721}.
  \href{https://doi.org/10.1007/11880240_49}{Springer Berlin Heidelberg},
  \href{https://doi.org/10.1007/11880240_49}{2006}.
  \href{https://doi.org/10.1007/11880240_49}
{doi: \textsf{%
10\hspace{.1pt}\discretionary{.}{%
}{.}\hspace{.4pt}1007\discretionary{/}{%
}{/}11880240\_49}}


\bibitem{Gratzer-MathIntoLatex}
G.~Gr{\"a}tzer.
\newblock {\em Math into LaTeX}.
\newblock Birkh{\"a}user, 3rd ed., 2000.

\bibitem{Groce2004UnderstandingCounterexamplesExplain}
\href{https://doi.org/10.1007/978-3-540-27813-9_35}{A.~Groce, D.~Kroening, and
  F.~Lerda}.
\newblock \href{https://doi.org/10.1007/978-3-540-27813-9_35}{Understanding
  counterexamples with explain}.
\newblock \href{https://doi.org/10.1007/978-3-540-27813-9_35}{In {\em Computer
  Aided Verification}}, \href{https://doi.org/10.1007/978-3-540-27813-9_35}{pp.
  453--456}. \href{https://doi.org/10.1007/978-3-540-27813-9_35}{Springer
  Berlin Heidelberg},
  \href{https://doi.org/10.1007/978-3-540-27813-9_35}{2004}.
  \href{https://doi.org/10.1007/978-3-540-27813-9_35}
{doi: \textsf{%
10\hspace{.1pt}\discretionary{.}{%
}{.}\hspace{.4pt}1007\discretionary{/}{%
}{/}978\discretionary{%
}{-}{-}3\discretionary{%
}{-}{-}540\discretionary{%
}{-}{-}27813\discretionary{%
}{-}{-}9\_35}}


\bibitem{Groce2003WhatWentWrong}
\href{https://doi.org/10.1007/3-540-44829-2_8}{A.~Groce and W.~Visser}.
\newblock \href{https://doi.org/10.1007/3-540-44829-2_8}{What went wrong:
  Explaining counterexamples}.
\newblock \href{https://doi.org/10.1007/3-540-44829-2_8}{In {\em Model Checking
  Software}}, \href{https://doi.org/10.1007/3-540-44829-2_8}{pp. 121--136}.
  \href{https://doi.org/10.1007/3-540-44829-2_8}{Springer Berlin Heidelberg},
  \href{https://doi.org/10.1007/3-540-44829-2_8}{2003}.
  \href{https://doi.org/10.1007/3-540-44829-2_8}
{doi: \textsf{%
10\hspace{.1pt}\discretionary{.}{%
}{.}\hspace{.4pt}1007\discretionary{/}{%
}{/}3\discretionary{%
}{-}{-}540\discretionary{%
}{-}{-}44829\discretionary{%
}{-}{-}2\_8}}


\bibitem{Gunnning2016Explainable}
D.~Gunning.
\newblock Explainable artificial intelligence ({XAI}).
\newblock Technical report, {Defense Advanced Research Projects Agency}
  ({DARPA}), 2016.

\bibitem{WEB:Haverbeke2011CodeMirror}
M.~Haverbeke.
\newblock {CodeMirror}, 2011.
\newblock \url{https://codemirror.net/}.

\bibitem{Hohman2019TeleGam}
\href{https://doi.org/10.1109/VISUAL.2019.8933695}{F.~{Hohman},
  A.~{Srinivasan}, and S.~M. {Drucker}}.
\newblock \href{https://doi.org/10.1109/VISUAL.2019.8933695}{{TeleGam}:
  Combining visualization and verbalization for interpretable machine
  learning}.
\newblock \href{https://doi.org/10.1109/VISUAL.2019.8933695}{In {\em IEEE
  Visualization Conference}},
  \href{https://doi.org/10.1109/VISUAL.2019.8933695}{pp. 151--155}.
  \href{https://doi.org/10.1109/VISUAL.2019.8933695}{{IEEE}},
  \href{https://doi.org/10.1109/VISUAL.2019.8933695}{2019}.
  \href{https://doi.org/10.1109/VISUAL.2019.8933695}
{doi: \textsf{%
10\hspace{.1pt}\discretionary{.}{%
}{.}\hspace{.4pt}1109\discretionary{/}{%
}{/}VISUAL\hspace{.1pt}\discretionary{.}{%
}{.}\hspace{.4pt}2019\hspace{.1pt}\discretionary{.}{%
}{.}\hspace{.4pt}8933695}}


\bibitem{IEEE-Verilog}
\href{https://doi.org/10.1109/IEEESTD.2006.99495}{{IEEE Computer Society}}.
\newblock \href{https://doi.org/10.1109/IEEESTD.2006.99495}{{\em {IEEE}
  Standard for {Verilog Hardware Description Language}}},
  \href{https://doi.org/10.1109/IEEESTD.2006.99495}{2006}.
  \href{https://doi.org/10.1109/IEEESTD.2006.99495}
{doi: \textsf{%
10\hspace{.1pt}\discretionary{.}{%
}{.}\hspace{.4pt}1109\discretionary{/}{%
}{/}IEEESTD\hspace{.1pt}\discretionary{.}{%
}{.}\hspace{.4pt}2006\hspace{.1pt}\discretionary{.}{%
}{.}\hspace{.4pt}99495}}


\bibitem{Jee2010FBDVerifierInteractiveVisual}
\href{https://search.informit.org/doi/10.3316/ielapa.448422067913862}{E.~Jee,
  S.~Jeon, S.~Cha, K.~Koh, J.~Yoo, G.~Park, and P.~Seong}.
\newblock
  \href{https://search.informit.org/doi/10.3316/ielapa.448422067913862}{Fbdverifier:
  Interactive and visual analysis of counter-example in formal verification of
  function block diagram}.
\newblock
  \href{https://search.informit.org/doi/10.3316/ielapa.448422067913862}{{\em
  Journal of Research and Practice in Information Technology}},
  \href{https://search.informit.org/doi/10.3316/ielapa.448422067913862}{42(3):171--–188},
  \href{https://search.informit.org/doi/10.3316/ielapa.448422067913862}{2010}.

\bibitem{Jeong2010VISAnalyzerVisual}
\href{https://doi.org/10.1109/isorc.2010.41}{S.~Jeong, J.~Yoo, and S.~Cha}.
\newblock \href{https://doi.org/10.1109/isorc.2010.41}{{VIS} analyzer: A visual
  assistant for {VIS} verification and analysis}.
\newblock \href{https://doi.org/10.1109/isorc.2010.41}{In {\em {IEEE}
  International Symposium on Object/Component/Service-Oriented Real-Time
  Distributed Computing}}.
  \href{https://doi.org/10.1109/isorc.2010.41}{{IEEE}},
  \href{https://doi.org/10.1109/isorc.2010.41}{Piscataway, NJ, USA},
  \href{https://doi.org/10.1109/isorc.2010.41}{2010}.
  \href{https://doi.org/10.1109/isorc.2010.41}
{doi: \textsf{%
10\hspace{.1pt}\discretionary{.}{%
}{.}\hspace{.4pt}1109\discretionary{/}{%
}{/}isorc\hspace{.1pt}\discretionary{.}{%
}{.}\hspace{.4pt}2010\hspace{.1pt}\discretionary{.}{%
}{.}\hspace{.4pt}41}}


\bibitem{DBLP:conf/cav/KaivolaGNTWPSTFRN09}
\href{https://doi.org/10.1007/978-3-642-02658-4_32}{R.~Kaivola, R.~Ghughal,
  N.~Narasimhan, A.~Telfer, J.~Whittemore, S.~Pandav, A.~Slobodov{\'a},
  C.~Taylor, V.~A. Frolov, E.~Reeber, and A.~Naik}.
\newblock \href{https://doi.org/10.1007/978-3-642-02658-4_32}{Replacing testing
  with formal verification in intel coretm i7 processor execution engine
  validation}.
\newblock \href{https://doi.org/10.1007/978-3-642-02658-4_32}{In {\em Intl.
  Conference on Computer Aided Verification}},
  \href{https://doi.org/10.1007/978-3-642-02658-4_32}{pp. 414--429}.
  \href{https://doi.org/10.1007/978-3-642-02658-4_32}{Springer},
  \href{https://doi.org/10.1007/978-3-642-02658-4_32}{2009}.
  \href{https://doi.org/10.1007/978-3-642-02658-4_32}
{doi: \textsf{%
10\hspace{.1pt}\discretionary{.}{%
}{.}\hspace{.4pt}1007\discretionary{/}{%
}{/}978\discretionary{%
}{-}{-}3\discretionary{%
}{-}{-}642\discretionary{%
}{-}{-}02658\discretionary{%
}{-}{-}4\_32}}


\bibitem{DBLP:conf/cav/KatzHIJLLSTWZDK19}
\href{https://doi.org/10.1007/978-3-030-25540-4_26}{G.~Katz, D.~A. Huang,
  D.~Ibeling, K.~Julian, C.~Lazarus, R.~Lim, P.~Shah, S.~Thakoor, H.~Wu,
  A.~Zeljic, D.~L. Dill, M.~J. Kochenderfer, and C.~W. Barrett}.
\newblock \href{https://doi.org/10.1007/978-3-030-25540-4_26}{The marabou
  framework for verification and analysis of deep neural networks}.
\newblock \href{https://doi.org/10.1007/978-3-030-25540-4_26}{In {\em Intl.
  Conference on Computer Aided Verification}},
  \href{https://doi.org/10.1007/978-3-030-25540-4_26}{pp. 443--452}.
  \href{https://doi.org/10.1007/978-3-030-25540-4_26}{Springer},
  \href{https://doi.org/10.1007/978-3-030-25540-4_26}{2019}.
  \href{https://doi.org/10.1007/978-3-030-25540-4_26}
{doi: \textsf{%
10\hspace{.1pt}\discretionary{.}{%
}{.}\hspace{.4pt}1007\discretionary{/}{%
}{/}978\discretionary{%
}{-}{-}3\discretionary{%
}{-}{-}030\discretionary{%
}{-}{-}25540\discretionary{%
}{-}{-}4\_26}}


\bibitem{Kocher2018spectre}
P.~Kocher, J.~Horn, A.~Fogh, , D.~Genkin, D.~Gruss, W.~Haas, M.~Hamburg,
  M.~Lipp, S.~Mangard, T.~Prescher, M.~Schwarz, and Y.~Yarom.
\newblock Spectre attacks: Exploiting speculative execution.
\newblock In {\em IEEE Symposium on Security and Privacy}, 2019.

\bibitem{Kohlhase08adaptationof}
M.~Kohlhase, C.~Lange, C.~Müller, N.~Müller, and F.~Rabe.
\newblock Adaptation of notations in living mathematical documents, 2008.

\bibitem{kohlhase2012semantics}
M.~Kohlhase and F.~Rabe.
\newblock Semantics of openmath and mathml 3.
\newblock {\em Mathematics in Computer Science}, 6(3):235--260, 2012.

\bibitem{DBLP:journals/corr/KoleiniCM13}
\href{http://arxiv.org/abs/1306.5678}{M.~Koleini, M.~R. Clarkson, and K.~K.
  Micinski}.
\newblock \href{http://arxiv.org/abs/1306.5678}{A temporal logic of security}.
\newblock \href{http://arxiv.org/abs/1306.5678}{{\em CoRR}},
  \href{http://arxiv.org/abs/1306.5678}{abs/1306.5678},
  \href{http://arxiv.org/abs/1306.5678}{2013}.

\bibitem{Koutsofios1992DrawingGraphsDot}
E.~Koutsofios and S.~C. North.
\newblock Drawing graphs with dot.
\newblock Technical report, {AT\&T} Bell Laboratories, 1992.

\bibitem{Lahtinen2012ModelCheckingMethodology}
J.~Lahtinen, T.~Launiainen, K.~Heljanko, and J.~Ropponen.
\newblock {\em Model Checking Methodology for Large Systems, Faults and
  Asynchronous Behaviour: SARANA 2011 Work Report}.
\newblock VTT Technical Research Centre of Finland, Finland, 2012.

\bibitem{LaoTebar2016ProposalFC}
J.~Lao-Tebar, F.~Alvaro, and D.~Marques.
\newblock Proposal for coexistence of mathematical handwritten and keyboard
  input in a wysiwyg expression editor.
\newblock In {\em FM4M/MathUI/ThEdu/DP/WIP@CIKM}, 2016.

\bibitem{Larsen201820YearsUPPAAL}
\href{https://doi.org/10.1007/978-3-030-03427-6_18}{K.~G. Larsen, F.~Lorber,
  and B.~Nielsen}.
\newblock \href{https://doi.org/10.1007/978-3-030-03427-6_18}{20 years of
  {UPPAAL} enabled industrial model-based validation and beyond}.
\newblock \href{https://doi.org/10.1007/978-3-030-03427-6_18}{In {\em Lecture
  Notes in Computer Science}},
  \href{https://doi.org/10.1007/978-3-030-03427-6_18}{pp. 212--229}.
  \href{https://doi.org/10.1007/978-3-030-03427-6_18}{Springer International
  Publishing}, \href{https://doi.org/10.1007/978-3-030-03427-6_18}{2018}.
  \href{https://doi.org/10.1007/978-3-030-03427-6_18}
{doi: \textsf{%
10\hspace{.1pt}\discretionary{.}{%
}{.}\hspace{.4pt}1007\discretionary{/}{%
}{/}978\discretionary{%
}{-}{-}3\discretionary{%
}{-}{-}030\discretionary{%
}{-}{-}03427\discretionary{%
}{-}{-}6\_18}}


\bibitem{Lipp2018meltdown}
M.~Lipp, M.~Schwarz, D.~Gruss, T.~Prescher, W.~Haas, A.~Fogh, J.~Horn,
  S.~Mangard, P.~Kocher, D.~Genkin, Y.~Yarom, and M.~Hamburg.
\newblock Meltdown: Reading kernel memory from user space.
\newblock In {\em {USENIX} Security Symposium}, 2018.

\bibitem{Liu2020AutoCaption}
\href{https://doi.org/10.1109/PacificVis48177.2020.1043}{C.~{Liu}, L.~{Xie},
  Y.~{Han}, D.~{Wei}, and X.~{Yuan}}.
\newblock
  \href{https://doi.org/10.1109/PacificVis48177.2020.1043}{{AutoCaption}: An
  approach to generate natural language description from visualization
  automatically}.
\newblock \href{https://doi.org/10.1109/PacificVis48177.2020.1043}{In {\em
  Proc. {IEEE} Pacific Symposium on Visualization}},
  \href{https://doi.org/10.1109/PacificVis48177.2020.1043}{pp. 191--195}.
  \href{https://doi.org/10.1109/PacificVis48177.2020.1043}{{IEEE}},
  \href{https://doi.org/10.1109/PacificVis48177.2020.1043}{2020}.
  \href{https://doi.org/10.1109/PacificVis48177.2020.1043}
{doi: \textsf{%
10\hspace{.1pt}\discretionary{.}{%
}{.}\hspace{.4pt}1109\discretionary{/}{%
}{/}PacificVis48177\hspace{.1pt}\discretionary{.}{%
}{.}\hspace{.4pt}2020\hspace{.1pt}\discretionary{.}{%
}{.}\hspace{.4pt}1043}}


\bibitem{Loer2006IntegratedFrameworkAnalysis}
\href{https://doi.org/10.1007/s10515-006-7999-y}{K.~Loer and M.~D. Harrison}.
\newblock \href{https://doi.org/10.1007/s10515-006-7999-y}{An integrated
  framework for the analysis of dependable interactive systems ({IFADIS}): Its
  tool support and evaluation}.
\newblock \href{https://doi.org/10.1007/s10515-006-7999-y}{{\em Automated
  Software Engineering}},
  \href{https://doi.org/10.1007/s10515-006-7999-y}{13(4):469--496},
  \href{https://doi.org/10.1007/s10515-006-7999-y}{May 2006}.
  \href{https://doi.org/10.1007/s10515-006-7999-y}
{doi: \textsf{%
10\hspace{.1pt}\discretionary{.}{%
}{.}\hspace{.4pt}1007\discretionary{/}{%
}{/}s10515\discretionary{%
}{-}{-}006\discretionary{%
}{-}{-}7999\discretionary{%
}{-}{-}y}}


\bibitem{MannaPnueli1996Safety}
Z.~Manna and A.~Pnueli.
\newblock {\em Temporal verification of reactive systems - safety}.
\newblock Springer, 1995.

\bibitem{MartinezMaldonado2020FromData}
\href{https://doi.org/10.1145/3313831.3376148}{R.~Martinez-Maldonado,
  V.~Echeverria, G.~Fernandez~Nieto, and S.~Buckingham~Shum}.
\newblock \href{https://doi.org/10.1145/3313831.3376148}{From data to insights:
  A layered storytelling approach for multimodal learning analytics}.
\newblock \href{https://doi.org/10.1145/3313831.3376148}{In {\em Proc. {CHI}
  Conference on Human Factors in Computing Systems}},
  \href{https://doi.org/10.1145/3313831.3376148}{pp. 1--–15}.
  \href{https://doi.org/10.1145/3313831.3376148}{{ACM}},
  \href{https://doi.org/10.1145/3313831.3376148}{New York, NY, USA},
  \href{https://doi.org/10.1145/3313831.3376148}{2020}.
  \href{https://doi.org/10.1145/3313831.3376148}
{doi: \textsf{%
10\hspace{.1pt}\discretionary{.}{%
}{.}\hspace{.4pt}1145\discretionary{/}{%
}{/}3313831\hspace{.1pt}\discretionary{.}{%
}{.}\hspace{.4pt}3376148}}


\bibitem{New-FSM-to-VHDL}
\href{https://doi.org/10.1109/SCORED.2017.8305431}{S.~H. {Masoumi}, S.~A.~R.
  {Al-Haddad}, and F.~Z. {Rokhani}}.
\newblock \href{https://doi.org/10.1109/SCORED.2017.8305431}{New tool for
  converting high-level representations of finite state machines to verilog
  hdl}.
\newblock \href{https://doi.org/10.1109/SCORED.2017.8305431}{In {\em IEEE
  Student Conference on Research and Development}},
  \href{https://doi.org/10.1109/SCORED.2017.8305431}{pp. 1--6},
  \href{https://doi.org/10.1109/SCORED.2017.8305431}{2017}.
  \href{https://doi.org/10.1109/SCORED.2017.8305431}
{doi: \textsf{%
10\hspace{.1pt}\discretionary{.}{%
}{.}\hspace{.4pt}1109\discretionary{/}{%
}{/}SCORED\hspace{.1pt}\discretionary{.}{%
}{.}\hspace{.4pt}2017\hspace{.1pt}\discretionary{.}{%
}{.}\hspace{.4pt}8305431}}


\bibitem{Mathisen2019InsideInsightsIntegratingData}
\href{https://doi.org/10.1111/cgf.13717}{A.~Mathisen, T.~Horak, C.~N. Klokmose,
  K.~Grønbæk, and N.~Elmqvist}.
\newblock \href{https://doi.org/10.1111/cgf.13717}{{InsideInsights: Integrating
  Data-Driven Reporting in Collaborative Visual Analytics}}.
\newblock \href{https://doi.org/10.1111/cgf.13717}{{\em Computer Graphics
  Forum}}, \href{https://doi.org/10.1111/cgf.13717}{38(3)},
  \href{https://doi.org/10.1111/cgf.13717}{June 2019}.
  \href{https://doi.org/10.1111/cgf.13717}
{doi: \textsf{%
10\hspace{.1pt}\discretionary{.}{%
}{.}\hspace{.4pt}1111\discretionary{/}{%
}{/}cgf\hspace{.1pt}\discretionary{.}{%
}{.}\hspace{.4pt}13717}}


\bibitem{Michael2019TeachingRigorousDistributed}
\href{https://doi.org/10.1145/3302424.3303947}{E.~Michael, D.~Woos,
  T.~Anderson, M.~D. Ernst, and Z.~Tatlock}.
\newblock \href{https://doi.org/10.1145/3302424.3303947}{Teaching rigorous
  distributed systems with efficient model checking}.
\newblock \href{https://doi.org/10.1145/3302424.3303947}{In {\em Proc.
  {EuroSys} Conference}}, \href{https://doi.org/10.1145/3302424.3303947}{pp.
  32:1--32:15}. \href{https://doi.org/10.1145/3302424.3303947}{{ACM}},
  \href{https://doi.org/10.1145/3302424.3303947}{New York, NY, USA},
  \href{https://doi.org/10.1145/3302424.3303947}{2019}.
  \href{https://doi.org/10.1145/3302424.3303947}
{doi: \textsf{%
10\hspace{.1pt}\discretionary{.}{%
}{.}\hspace{.4pt}1145\discretionary{/}{%
}{/}3302424\hspace{.1pt}\discretionary{.}{%
}{.}\hspace{.4pt}3303947}}


\bibitem{WEB:MyScript2017}
{MyScript}.
\newblock Handwriting recognition - {MyScript}, 2017.
\newblock \url{https://www.myscript.com/handwriting-recognition}.

\bibitem{Obeid2020Chart}
\href{https://www.aclweb.org/anthology/2020.inlg-1.20}{J.~Obeid and E.~Hoque}.
\newblock
  \href{https://www.aclweb.org/anthology/2020.inlg-1.20}{{Chart-to-Text}:
  Generating natural language descriptions for charts by adapting the
  transformer model}.
\newblock \href{https://www.aclweb.org/anthology/2020.inlg-1.20}{In {\em Proc.
  International Conference on Natural Language Generation}},
  \href{https://www.aclweb.org/anthology/2020.inlg-1.20}{pp. 138--147}.
  \href{https://www.aclweb.org/anthology/2020.inlg-1.20}{Association for
  Computational Linguistics},
  \href{https://www.aclweb.org/anthology/2020.inlg-1.20}{Dublin, Ireland},
  \href{https://www.aclweb.org/anthology/2020.inlg-1.20}{2020}.

\bibitem{WEB:OpenCores}
{OpenCores.org}.
\newblock {OpenCores}, 1999.
\newblock \url{https://opencores.org/}.

\bibitem{WEB:OpenJSFoundation2009Node.js}
{OpenJS Foundation}.
\newblock Node.js, 2009.
\newblock \url{http://nodejs.org/}.

\bibitem{Pakonen2018CounterexampleVisualizationExplanation}
\href{https://doi.org/10.1109/indin.2018.8472025}{A.~Pakonen, I.~Buzhinsky, and
  V.~Vyatkin}.
\newblock \href{https://doi.org/10.1109/indin.2018.8472025}{Counterexample
  visualization and explanation for function block diagrams}.
\newblock \href{https://doi.org/10.1109/indin.2018.8472025}{In {\em {IEEE}
  International Conference on Industrial Informatics}},
  \href{https://doi.org/10.1109/indin.2018.8472025}{pp. 747--753}.
  \href{https://doi.org/10.1109/indin.2018.8472025}{{IEEE}},
  \href{https://doi.org/10.1109/indin.2018.8472025}{2018}.
  \href{https://doi.org/10.1109/indin.2018.8472025}
{doi: \textsf{%
10\hspace{.1pt}\discretionary{.}{%
}{.}\hspace{.4pt}1109\discretionary{/}{%
}{/}indin\hspace{.1pt}\discretionary{.}{%
}{.}\hspace{.4pt}2018\hspace{.1pt}\discretionary{.}{%
}{.}\hspace{.4pt}8472025}}


\bibitem{Patil2015CounterexampleguidedSimulation}
\href{https://doi.org/10.1109/indin.2015.7281905}{S.~Patil, V.~Vyatkin, and
  C.~Pang}.
\newblock
  \href{https://doi.org/10.1109/indin.2015.7281905}{Counterexample-guided
  simulation framework for formal verification of flexible automation systems}.
\newblock \href{https://doi.org/10.1109/indin.2015.7281905}{In {\em {IEEE}
  International Conference on Industrial Informatics}}.
  \href{https://doi.org/10.1109/indin.2015.7281905}{{IEEE}},
  \href{https://doi.org/10.1109/indin.2015.7281905}{Piscataway, NJ, USA},
  \href{https://doi.org/10.1109/indin.2015.7281905}{2015}.
  \href{https://doi.org/10.1109/indin.2015.7281905}
{doi: \textsf{%
10\hspace{.1pt}\discretionary{.}{%
}{.}\hspace{.4pt}1109\discretionary{/}{%
}{/}indin\hspace{.1pt}\discretionary{.}{%
}{.}\hspace{.4pt}2015\hspace{.1pt}\discretionary{.}{%
}{.}\hspace{.4pt}7281905}}


\bibitem{HDL-FSM}
V.~A. Pedroni.
\newblock {\em Finite State Machines in Hardware: Theory and Design (with VHDL
  and SystemVerilog)}.
\newblock The MIT Press, 2013.

\bibitem{Pollanen2014TowardsAU}
M.~Pollanen, J.~Hooper, B.~Cater, and S.~Kang.
\newblock Towards a universal interface for real-time mathematical
  communication.
\newblock In {\em CICM Workshops}, 2014.

\bibitem{Roberts2007STAR_CMV}
\href{https://doi.org/10.1109/cmv.2007.20}{J.~C. Roberts}.
\newblock \href{https://doi.org/10.1109/cmv.2007.20}{State of the art:
  Coordinated \& multiple views in exploratory visualization}.
\newblock \href{https://doi.org/10.1109/cmv.2007.20}{In {\em Proc. {IEEE}
  Conference on Coordinated and Multiple Views in Exploratory Visualization}},
  \href{https://doi.org/10.1109/cmv.2007.20}{pp. 61--71}.
  \href{https://doi.org/10.1109/cmv.2007.20}{{IEEE}},
  \href{https://doi.org/10.1109/cmv.2007.20}{Piscataway, NJ, USA},
  \href{https://doi.org/10.1109/cmv.2007.20}{July 2007}.
  \href{https://doi.org/10.1109/cmv.2007.20}
{doi: \textsf{%
10\hspace{.1pt}\discretionary{.}{%
}{.}\hspace{.4pt}1109\discretionary{/}{%
}{/}cmv\hspace{.1pt}\discretionary{.}{%
}{.}\hspace{.4pt}2007\hspace{.1pt}\discretionary{.}{%
}{.}\hspace{.4pt}20}}


\bibitem{Romat2021ExpressiveAuthoringNode}
\href{https://doi.org/10.1109/tvcg.2019.2950932}{H.~Romat, C.~Appert, and
  E.~Pietriga}.
\newblock \href{https://doi.org/10.1109/tvcg.2019.2950932}{Expressive authoring
  of node-link diagrams with graphies}.
\newblock \href{https://doi.org/10.1109/tvcg.2019.2950932}{{\em {{IEEE} Trans.
  Visualization and Computer Graphics}}},
  \href{https://doi.org/10.1109/tvcg.2019.2950932}{27(4):2329--2340},
  \href{https://doi.org/10.1109/tvcg.2019.2950932}{Apr. 2021}.
  \href{https://doi.org/10.1109/tvcg.2019.2950932}
{doi: \textsf{%
10\hspace{.1pt}\discretionary{.}{%
}{.}\hspace{.4pt}1109\discretionary{/}{%
}{/}tvcg\hspace{.1pt}\discretionary{.}{%
}{.}\hspace{.4pt}2019\hspace{.1pt}\discretionary{.}{%
}{.}\hspace{.4pt}2950932}}


\bibitem{Rothenberger2016IntegrationAnalysisAlternative}
\href{https://doi.org/10.3929/ETHZ-A-010608394}{F.~Rothenberger}.
\newblock \href{https://doi.org/10.3929/ETHZ-A-010608394}{Integration and
  analysis of alternative smt solvers for software verification}.
\newblock \href{https://doi.org/10.3929/ETHZ-A-010608394}{Master's thesis},
  \href{https://doi.org/10.3929/ETHZ-A-010608394}{ETH Zurich},
  \href{https://doi.org/10.3929/ETHZ-A-010608394}{2016}.
  \href{https://doi.org/10.3929/ETHZ-A-010608394}
{doi: \textsf{%
10\hspace{.1pt}\discretionary{.}{%
}{.}\hspace{.4pt}3929\discretionary{/}{%
}{/}ETHZ\discretionary{%
}{-}{-}A\discretionary{%
}{-}{-}010608394}}


\bibitem{Schuppan2005ShortestCounterexamplesSymbolic}
\href{https://doi.org/10.1007/978-3-540-31980-1_32}{V.~Schuppan and A.~Biere}.
\newblock \href{https://doi.org/10.1007/978-3-540-31980-1_32}{Shortest
  counterexamples for symbolic model checking of {LTL} with past}.
\newblock \href{https://doi.org/10.1007/978-3-540-31980-1_32}{In {\em Tools and
  Algorithms for the Construction and Analysis of Systems}},
  \href{https://doi.org/10.1007/978-3-540-31980-1_32}{pp. 493--509}.
  \href{https://doi.org/10.1007/978-3-540-31980-1_32}{Springer Berlin
  Heidelberg}, \href{https://doi.org/10.1007/978-3-540-31980-1_32}{2005}.
  \href{https://doi.org/10.1007/978-3-540-31980-1_32}
{doi: \textsf{%
10\hspace{.1pt}\discretionary{.}{%
}{.}\hspace{.4pt}1007\discretionary{/}{%
}{/}978\discretionary{%
}{-}{-}3\discretionary{%
}{-}{-}540\discretionary{%
}{-}{-}31980\discretionary{%
}{-}{-}1\_32}}


\bibitem{WEB:SeoulOh2012MathQuill}
H.~Seoul-Oh, J.~Adkisson, and M.~Stufflebeam.
\newblock Mathquill, 2012.
\newblock \url{http://mathquill.com/}.

\bibitem{Sevastjanova2018Going}
R.~Sevastjanova, F.~Beck, B.~Ell, C.~Turkay, R.~Henkin, M.~Butt, D.~A. Keim,
  and M.~El-Assady.
\newblock Going beyond visualization: Verbalization as complementary medium to
  explain machine learning models.
\newblock In {\em Workshop on Visualization for AI Explainability at IEEE VIS},
  2018.

\bibitem{Shi2021Calliope}
\href{https://doi.org/10.1109/TVCG.2020.3030403}{D.~{Shi}, X.~{Xu}, F.~{Sun},
  Y.~{Shi}, and N.~{Cao}}.
\newblock \href{https://doi.org/10.1109/TVCG.2020.3030403}{Calliope: Automatic
  visual data story generation from a spreadsheet}.
\newblock \href{https://doi.org/10.1109/TVCG.2020.3030403}{{\em {{IEEE} Trans.
  Visualization and Computer Graphics}}},
  \href{https://doi.org/10.1109/TVCG.2020.3030403}{27(2):453--463},
  \href{https://doi.org/10.1109/TVCG.2020.3030403}{2021}.
  \href{https://doi.org/10.1109/TVCG.2020.3030403}
{doi: \textsf{%
10\hspace{.1pt}\discretionary{.}{%
}{.}\hspace{.4pt}1109\discretionary{/}{%
}{/}TVCG\hspace{.1pt}\discretionary{.}{%
}{.}\hspace{.4pt}2020\hspace{.1pt}\discretionary{.}{%
}{.}\hspace{.4pt}3030403}}


\bibitem{Spinner2020explAIner}
\href{https://doi.org/10.1109/TVCG.2019.2934629}{T.~{Spinner}, U.~{Schlegel},
  H.~{Schäfer}, and M.~{El-Assady}}.
\newblock \href{https://doi.org/10.1109/TVCG.2019.2934629}{{explAIner}: A
  visual analytics framework for interactive and explainable machine learning}.
\newblock \href{https://doi.org/10.1109/TVCG.2019.2934629}{{\em {{IEEE} Trans.
  Visualization and Computer Graphics}}},
  \href{https://doi.org/10.1109/TVCG.2019.2934629}{26(1):1064--1074},
  \href{https://doi.org/10.1109/TVCG.2019.2934629}{2020}.
  \href{https://doi.org/10.1109/TVCG.2019.2934629}
{doi: \textsf{%
10\hspace{.1pt}\discretionary{.}{%
}{.}\hspace{.4pt}1109\discretionary{/}{%
}{/}TVCG\hspace{.1pt}\discretionary{.}{%
}{.}\hspace{.4pt}2019\hspace{.1pt}\discretionary{.}{%
}{.}\hspace{.4pt}2934629}}


\bibitem{Spreafico2020NeuralDataDriven}
\href{https://doi.org/10.1145/3399715.3399829}{A.~Spreafico and G.~Carenini}.
\newblock \href{https://doi.org/10.1145/3399715.3399829}{Neural data-driven
  captioning of time-series line charts}.
\newblock \href{https://doi.org/10.1145/3399715.3399829}{In {\em Proc.
  International Conference on Advanced Visual Interfaces}}.
  \href{https://doi.org/10.1145/3399715.3399829}{{ACM}},
  \href{https://doi.org/10.1145/3399715.3399829}{New York, NY, USA},
  \href{https://doi.org/10.1145/3399715.3399829}{2020}.
  \href{https://doi.org/10.1145/3399715.3399829}
{doi: \textsf{%
10\hspace{.1pt}\discretionary{.}{%
}{.}\hspace{.4pt}1145\discretionary{/}{%
}{/}3399715\hspace{.1pt}\discretionary{.}{%
}{.}\hspace{.4pt}3399829}}


\bibitem{Srinivasan2019AugmentingVisualizationsInteractive}
\href{https://doi.org/10.1109/tvcg.2018.2865145}{A.~Srinivasan, S.~M. Drucker,
  A.~Endert, and J.~Stasko}.
\newblock \href{https://doi.org/10.1109/tvcg.2018.2865145}{Augmenting
  visualizations with interactive data facts to facilitate interpretation and
  communication}.
\newblock \href{https://doi.org/10.1109/tvcg.2018.2865145}{{\em {{IEEE} Trans.
  Visualization and Computer Graphics}}},
  \href{https://doi.org/10.1109/tvcg.2018.2865145}{25(1):672--681},
  \href{https://doi.org/10.1109/tvcg.2018.2865145}{Jan. 2019}.
  \href{https://doi.org/10.1109/tvcg.2018.2865145}
{doi: \textsf{%
10\hspace{.1pt}\discretionary{.}{%
}{.}\hspace{.4pt}1109\discretionary{/}{%
}{/}tvcg\hspace{.1pt}\discretionary{.}{%
}{.}\hspace{.4pt}2018\hspace{.1pt}\discretionary{.}{%
}{.}\hspace{.4pt}2865145}}


\bibitem{Strobelt2019Seq2seq}
\href{https://doi.org/10.1109/TVCG.2018.2865044}{H.~{Strobelt}, S.~{Gehrmann},
  M.~{Behrisch}, A.~{Perer}, H.~{Pfister}, and A.~M. {Rush}}.
\newblock \href{https://doi.org/10.1109/TVCG.2018.2865044}{{Seq2seq-Vis}: A
  visual debugging tool for sequence-to-sequence models}.
\newblock \href{https://doi.org/10.1109/TVCG.2018.2865044}{{\em {{IEEE} Trans.
  Visualization and Computer Graphics}}},
  \href{https://doi.org/10.1109/TVCG.2018.2865044}{25(1):353--363},
  \href{https://doi.org/10.1109/TVCG.2018.2865044}{2019}.
  \href{https://doi.org/10.1109/TVCG.2018.2865044}
{doi: \textsf{%
10\hspace{.1pt}\discretionary{.}{%
}{.}\hspace{.4pt}1109\discretionary{/}{%
}{/}TVCG\hspace{.1pt}\discretionary{.}{%
}{.}\hspace{.4pt}2018\hspace{.1pt}\discretionary{.}{%
}{.}\hspace{.4pt}2865044}}


\bibitem{I2C-article}
W.~Thielke.
\newblock {Code geknackt}.
\newblock
  \url{https://www.focus.de/finanzen/news/krankenkassen-code-geknackt_aid_148829.html},
  2013.

\bibitem{Wiehr2020SafeHandoverMixed}
F.~Wiehr, A.~Hirsch, F.~Daiber, A.~Kruger, A.~Kovtunova, S.~Borgwardt,
  E.~Chang, V.~Demberg, M.~Steinmetz, and H.~Jorg.
\newblock Safe handover in mixed-initiative control for cyber-physical systems,
  2020.

\bibitem{WigmHuntPflu2009dp}
\href{https://www.learntechlib.org/p/30301}{A.~Wigmore, G.~Hunter, E.~Pflügel,
  J.~Denholm-Price, and V.~Binelli}.
\newblock \href{https://www.learntechlib.org/p/30301}{Using automatic speech
  recognition to dictate mathematical expressions: The development of the
  “talkmaths” application at kingston university}.
\newblock \href{https://www.learntechlib.org/p/30301}{{\em Journal of Computers
  in Mathematics and Science Teaching}},
  \href{https://www.learntechlib.org/p/30301}{28(2):177--189},
  \href{https://www.learntechlib.org/p/30301}{Apr. 2009}.

\bibitem{Wolfram-SMC}
\href{https://doi.org/10.1145/3341.3347}{S.~Wolfram}.
\newblock \href{https://doi.org/10.1145/3341.3347}{Symbolic mathematical
  computation}.
\newblock \href{https://doi.org/10.1145/3341.3347}{{\em Communications of the
  ACM}}, \href{https://doi.org/10.1145/3341.3347}{28(4):390--–394},
  \href{https://doi.org/10.1145/3341.3347}{Apr. 1985}.
  \href{https://doi.org/10.1145/3341.3347}
{doi: \textsf{%
10\hspace{.1pt}\discretionary{.}{%
}{.}\hspace{.4pt}1145\discretionary{/}{%
}{/}3341\hspace{.1pt}\discretionary{.}{%
}{.}\hspace{.4pt}3347}}


\bibitem{Woos2019StepDebuggerDistributed}
D.~Woos.
\newblock {\em A Step-through Debugger for Distributed Systems}.
\newblock PhD thesis, University of Washington, 2019.

\bibitem{Xu2020SurveyAnalysisUser}
\href{https://doi.org/10.1111/cgf.14035}{K.~Xu, A.~Ottley, C.~Walchshofer,
  M.~Streit, R.~Chang, and J.~Wenskovitch}.
\newblock \href{https://doi.org/10.1111/cgf.14035}{Survey on the analysis of
  user interactions and visualization provenance}.
\newblock \href{https://doi.org/10.1111/cgf.14035}{{\em Computer Graphics
  Forum}}, \href{https://doi.org/10.1111/cgf.14035}{39(3):757--783},
  \href{https://doi.org/10.1111/cgf.14035}{June 2020}.
  \href{https://doi.org/10.1111/cgf.14035}
{doi: \textsf{%
10\hspace{.1pt}\discretionary{.}{%
}{.}\hspace{.4pt}1111\discretionary{/}{%
}{/}cgf\hspace{.1pt}\discretionary{.}{%
}{.}\hspace{.4pt}14035}}


\end{thebibliography}
\end{document}